\documentstyle[12pt,epsf]{article} 

\begin{document} 
  
\centerline{\large\bf Center vortex model for the infrared sector of}
\vspace{0.2cm}
\centerline{\large\bf Yang-Mills theory -- Quenched Dirac spectrum}
\vspace{0.2cm}
\centerline{\large\bf and chiral condensate}

\bigskip
\centerline{M.~Engelhardt\footnote{Supported by DFG 
under grants Re 856/4-1 and Al 279/3-3.} }
\vspace{0.2 true cm} 
\centerline{\em Institut f\"ur Theoretische Physik, Universit\"at 
T\"ubingen }
\centerline{\em D--72076 T\"ubingen, Germany}
  
\begin{abstract}
The Dirac operator describing the coupling of continuum quark fields to
$SU(2)$ center vortex world-surfaces composed of elementary squares on a
hypercubic lattice is constructed. It is used to evaluate the quenched 
Dirac spectral density in the random vortex world-surface model, which
previously has been shown to quantitatively reproduce both the confinement
properties and the topological susceptibility of $SU(2)$ Yang-Mills theory.
Under certain conditions on the modeling of the vortex gauge field,
a behavior of the quenched chiral condensate as a function of temperature
is obtained which is consistent with measurements in $SU(2)$ lattice
Yang-Mills theory.
\end{abstract}

\vskip .5truecm
\noindent
PACS: 12.38.Aw, 12.38.Mh, 12.40.-y

\noindent
Keywords: Center vortices, infrared effective theory,
Dirac spectrum, spontaneous chiral symmetry breaking
\medskip

\section{Introduction}
The random vortex surface model \cite{selprep},\cite{preptop}
is an infrared effective theory which aims to describe the
relevant physical fluctuations determining the low-energy nonperturbative
sector of the strong interaction. It is based on collective magnetic vortex
degrees of freedom \cite{thoo}-\cite{tomfe} which are represented by closed
two-dimensional world-surfaces in four-dimensional (Euclidean) space-time.
The chromomagnetic flux carried by the vortices is quantized according
to the center of the gauge group. The vortex world-surfaces are treated
as random surfaces, an ensemble of which in practice is generated
using Monte Carlo methods on a hypercubic lattice; the surfaces are
composed of elementary squares (plaquettes) on that lattice. The spacing
of the lattice is a fixed physical quantity, related to an intrinsic
thickness of the vortex fluxes, and represents the ultraviolet cutoff
inherent in any infrared effective framework. This model
can be adjusted such as to quantitatively reproduce the confinement
properties of $SU(2)$ lattice gauge theory \cite{selprep}, including
the finite-temperature transition to the deconfined phase. It 
simultaneously predicts a quantitatively correct topological susceptibility,
again as a function of temperature \cite{preptop}. The properties of the
model vortices closely parallel the ones obtained for vortex
configurations extracted from the full $SU(2)$ lattice Yang-Mills ensemble
using an appropriate gauge fixing and projection procedure
\cite{deb97},\cite{giedt}, cf.~\cite{temp}-\cite{bertop}.

The aim of the present work is to investigate whether also the third
central nonperturbative phenomenon determining low-energy strong
interaction physics, namely the spontaneous breaking of chiral symmetry,
can be correctly described within the random vortex surface model.
For this purpose, a number of technical developments are carried out. The
coupling of the gluonic vortex degrees of freedom to (continuum) quark
fields is governed by the Dirac operator; in order to construct this
operator, an explicit representation of vortex configurations in terms of
continuum gauge fields is needed. Having specified such a representation,
matrix elements of the Dirac operator can be calculated in a truncated
infrared quark basis. Via the (ensemble-averaged) spectral distribution of
the resulting Dirac matrix, which is obtained numerically, the (quenched)
chiral condensate is extracted; this condensate can be used as an order
parameter for spontaneous chiral symmetry breaking. Under certain conditions
on the explicit vortex gauge field modeling, the chiral condensate as a
function of temperature behaves in a manner consistent with $SU(2)$ lattice
gauge theory.

To date, several explorations of spontaneous chiral symmetry breaking
within the vortex picture have been presented. One of these lines
of investigation \cite{forc1}-\cite{forc3} deals with the vortex
configurations extracted from the full $SU(2)$ lattice Yang-Mills ensemble
by gauge fixing and projection \cite{deb97},\cite{giedt}. Removing these
vortices from $SU(2)$ lattice configurations leads to the restoration of
chiral symmetry \cite{forc1}; on the other hand, the extracted vortices
by themselves break chiral symmetry at zero temperature \cite{forc2}. In
the chiral limit, a divergence of the chiral condensate is observed which
is reminiscent of the behavior detected in full $SU(2)$ Yang-Mills theory
using domain-wall fermions \cite{chen}; such a divergence will also be
observed in the Dirac spectra obtained in the present work,
cf.~section \ref{ressec}. Crossing into the deconfined regime,
chiral symmetry remains broken, to an extent which differs
qualitatively \cite{forc3} from the breaking found in the
aforementioned domain-wall fermion calculations. In view of this result,
the significance of the chiral condensate induced by vortices and its
relation to the condensate obtained in full Yang-Mills theory remained in
doubt. In the present work, the behavior of the chiral condensate in the
high-temperature regime will turn out to depend strongly on the modeling
of the vortex gauge field. Only a sufficiently smooth definition of the
gauge field yields a behavior of the chiral condensate consistent with
full Yang-Mills theory. Possibly the vortex extraction procedure used
in \cite{forc3} endows the vortices with excessive spurious ultraviolet
disorder which leads to the unnatural result for the chiral condensate
in the deconfined phase.

On the other hand, quite recently an effective vortex model (in the guise
of a $Z(2)$ gauge theory in the strong coupling limit) was considered
\cite{tomcsb} which to a certain extent follows the logic of the
random vortex surface model investigated here. At the expense of a more
limited scope of applicability (the ultraviolet cutoff is e.g.~too low
to permit a description of the deconfinement phase transition), the
dynamics is simplified to the point where analytical estimates of the
chiral condensate become possible. These estimates appear to be close
to phenomenologically accepted values {\em for the unquenched, dynamical
quark case}. However, the extreme limit considered seems rather artificial
in that the quenched and the dynamical quark cases yield comparable chiral
condensates. Presumably, since one possibility of arriving at estimates
of the type used in \cite{tomcsb} appears to lie in invoking the
large-$N_c$ expansion ($N_c$ denoting the number of colors), this should not
come as a surprise; the large-$N_c$ expansion is by construction quenched to
leading order, dynamical quark effects only entering subdominant terms.
By contrast, the quenched chiral condensate obtained in lattice Yang-Mills
theory \cite{hatep} exceeds the one expected phenomenologically in the
presence of dynamical quarks by about an order of magnitude; clearly,
the quark determinant weighting of the dynamical quark ensemble must be
viewed as one of the dominant physical effects in order to reconcile the
quenched approximation with the dynamical quark case. In view of this,
the meaning of the chiral condensate discussed in \cite{tomcsb} remains
unclear, as does the question whether the quoted values in fact support
or weaken the basis for that model.

\section{General remarks on the formulation of a vortex gauge field model}
\label{gensec}
In the applications of the random vortex surface model hitherto
considered \cite{selprep},\cite{preptop}, it was not necessary to
explicitly construct a gauge field associated the vortex configurations
arising in the model. To evaluate confinement properties, encoded in
Wilson loops \cite{selprep}, it is sufficient to specify the space-time
location of the vortex world-surfaces; to evaluate the topological 
charge \cite{preptop}, it is sufficient to specify the vortex field
strength, which is localized on those surfaces. By contrast,
construction of the Dirac operator implies specifying the gauge field
itself. The freedom inherent in this specification entails
ambiguities\footnote{Further to the ambiguity discussed in the main text,
an additional one in principle can arise from the fact that, given a
particular (vortex) field strength, in the full space of nonabelian gauge
fields, this field strength may be encoded by several gauge-nonequivalent
gauge fields \cite{wuy}. The latter however will in general yield different
Dirac spectra.} due to the fact that the concept of an infrared
effective model of gauge and quark fields to a certain extent clashes
with gauge invariance; after all, the distinction whether a given
configuration is infrared or not is a gauge-dependent one. In an
infrared effective model of gauge and quark fields, manifest gauge
invariance can only be sustained with respect to
gauge transformations $U$ which are themselves infrared. Specifically,
when using a truncated basis of quark wave functions (as will be done
in this investigation), invariance under a change of the gauge in which
the gauge field ensemble is given is only guaranteed to the extent that,
for a quark wave function $q$ represented in the aforementioned basis,
also $Uq$ can be represented to good accuracy in that basis. Conceptually,
this is not an obstacle to model-building; it just implies that a full
specification of the model must include a fixed choice of gauge, and
that predictive power is curtailed to the extent that observables
depend on the precise gauge field modeling. In general, some
observables have to be used to fix the model before additional ones
can be predicted.

Ideally, the most consistent gauge field description of a given vortex
field strength ensemble would be one in which the gauge field is itself
as smooth, i.e. infrared, as possible, such as the Landau gauge.
In practice, the choice is constrained by considerations of technical
manageability, and also by certain idealizations of the vortex
configurations which already occur on the level of the field strengths,
cf.~below. In order to obtain information about the dependence of observables
on the construction of the gauge field, in this work different choices will
be explored, which, configuration by configuration, describe the same vortex
world-surfaces, but are related by gauge transformations which by no means
are infrared. It will turn out that chiral symmetry breaking in the confined
phase is quite robust as the modeling of the gauge field is changed;
on the other hand, the Dirac spectrum in the deconfined, high
temperature phase exhibits a substantial dependence on the gauge field
modeling - gratifyingly, the most consistent (smoothest) construction
turns out to best reproduce the results of full Yang-Mills measurements.

Before proceeding to present the construction of the gauge field in detail,
a few remarks are in order concerning the idealizations of the vortex
field strengths mentioned further above, which were already introduced
in previous considerations of the random vortex surface model.
The vortices hitherto used {\em to evaluate observables} 
within the model were idealized in two ways: For one, vortex fluxes were
taken to be infinitely thin, i.e. the associated world-surfaces are
truly two-dimensional, with no transverse thickness in the remaining
two space-time directions. Note that this idealization is 
gauge-invariant; it is a statement about the structure of the
modulus of the field strength tr$\, F^2 $. On the other hand, as a vector
in color space, the field strength $F$ on vortex surfaces was
taken to point exclusively into the positive or the negative 3-direction;
generic vortex surfaces are non-oriented, i.e. consist of patches of
alternating orientation. Note that the color direction of $F$ can be
changed by gauge transformations; thus, the configurations hitherto
used can be viewed as having been specified in a gauge of the
maximally Abelian type.

It should be emphasized that these are the properties of the field
strengths used {\em to evaluate observables} within the model.
By contrast, in the generation of the vortex surface ensemble,
a physical thickness of the vortices is indeed implicit \cite{selprep}:
A {\em finite} action density is associated with vortex surface area and
curvature; infinitely thin vortex fluxes of course formally would carry
divergent tr$\, F^2 $. Also, vortex surfaces cannot be packed arbitrarily
densely within the model; instead, there is a fixed ultraviolet cutoff,
interpreted as a consequence of the vortex thickness. The construction
and interpretation of the vortex dynamics is discussed at length
in \cite{selprep}.

The idealized thin vortex configurations described above are, however,
entirely sufficient for an unambiguous evaluation of the observables 
considered to date. Neither the asymptotic string tension 
\cite{selprep},\cite{casscal} nor the topological charge 
\cite{preptop},\cite{cont}-\cite{reineu} depend on any assumption about
the transverse structure of the vortex fluxes, i.e.~whether they are 
thickened or not. Also, both observables are virtually independent
of assumptions about the color orientation of the vortex field strength.
For the topological charge, this is somewhat less obvious than for
the Wilson loop; this issue is discussed in detail in \cite{preptop}.

On the other hand, in the case of the Dirac operator spectrum to be 
discussed here, the issue is a priori not as clear-cut; it is therefore
necessary to be specific about how the aforementioned idealizations will
be dealt with. In the following treatment, the idealization of thinness of
vortex flux will be maintained; a sufficiently simple and manageable
construction of manifestly thick vortex fluxes did not become apparent
in the course of this investigation\footnote{It should be noted that 
the thickness of realistic vortex fluxes (about 1 fm in 
diameter \cite{deb97},\cite{giedt}) noticeably exceeds the typical
distance separating neighboring vortices; a two-dimensional plane in
space-time is, on the average, pierced by 1.8 vortices/fm$^2 $ at zero
temperature \cite{selprep}. Physical vortex fluxes overlap to a
considerable extent.}. Instead, a model assumption will
be used which is familiar e.g. from the Nambu--Jona-Lasinio (NJL) quark
model. Namely, as already hinted at above, the quark modes will be subject
to a fixed ultraviolet cutoff, the same one which is implicit in the model 
vortex dynamics via the spacing of the lattice on which the vortex surfaces
are defined \cite{selprep}. In the NJL model, the quark current-current 
interaction formally is point-like; however, the ultraviolet cutoff on the
quark modes effectively smears out the four-quark vertex into a nonlocal
interaction. In complete analogy, within the present treatment, the 
interaction between a quark and a vortex formally takes place on an
infinitely thin submanifold (to be specific, a three-dimensional volume,
the two-dimensional boundary of which precisely gives the location of
the vortex field strength, cf.~section \ref{locsec}). However, the
ultraviolet cutoff on the quark modes will effectively smear this out in
the direction perpendicular to the volume (and, at its boundary, in the
two directions perpendicular to the idealized thin vortex field strength).
Thus, the thickening of the vortices is not carried out explicitly in
the vortex gauge fields, but is effectively accomplished by smearing out
the vortex-quark vertex via the ultraviolet cutoff on the quark fields.

Turning to the issue of the color orientation of the vortex field
strength, a more sophisticated approach will be pursued, for the 
following reason. If one works with a globally defined field strength,
pointing either into the positive or the negative 3-direction in color space,
and one attempts to globally define an Abelian gauge field generating
this field strength, then, in the (generic) case of nonorientable vortex
surfaces, one necessarily introduces Dirac string
world-surfaces into the description, emanating from the lines on the
vortex surfaces at which the orientation switches. After all, a globally
defined (Abelian) gauge field must satisfy continuity of (Abelian)
magnetic flux. This would not constitute a problem if one worked with
a complete quark basis, i.e.~solved the Dirac equation exactly. Then,
the quark wave functions would exhibit singularities along the Dirac
strings which would serve to cancel any physical effect of these strings,
i.e.~the Dirac strings would be unobservable, as they should be.
However, in any truncated calculation such as will be pursued here,
the cancellation would not be perfect; instead, the Dirac strings
effectively would act as additional physical magnetic fluxes (of double
magnitude compared with vortex fluxes) with which the quarks can interact.
Thus, more magnetic disorder would be effectively present than the
model aims to describe.

The way out of this dilemma lies in instead using a Wu-Yang construction
of the gauge field \cite{wuy}. In other words, the 
gauge field will be defined on local patches, which are then related to one
another via transition functions on their overlaps; the transition functions
in general will have to be non-Abelian. The patches should be chosen
sufficiently small, such that, on any given patch, the vortex surfaces
can be oriented; in this case, no Dirac strings arise in the associated
Abelian gauge field, which is then well-defined on the whole patch.
The reader should however be forewarned that constraints on the
space-time form of the transition functions in generic vortex
configurations will force the reintroduction of some spurious fluxes
(via additional non-Abelian gauge field components) which are formally
similar to Dirac strings, but with the {\em crucial difference that they
manifestly decouple} from the smooth infrared quark fields. Thus,
spurious fluxes are not completely eliminated by the construction
presented below; rather, they are recast in an innocuous form.
In practice, the Dirac equation will be solved using the finite
element method, i.e. using a basis of localized quark wave functions.
Thus, it is natural (and sufficient) to use the supports of the individual
basis functions, i.e. the finite elements, as the space-time patches
on which the gauge field is locally defined. Specifics on these finite
elements follow further below.

\pagebreak

\section{Construction of the vortex gauge field}
\label{gfsec}
\subsection{Local structure of the gauge field}
\label{locsec}
A closed $SU(2)$ vortex world-surface $S$ along with its quantized
chromomagnetic flux can, to a large extent, be characterized by the
gauge-invariant property
\begin{equation}
W [C] = (-1)^{L(C,S)}
\label{lino}
\end{equation}
for any Wilson loop $W[C]$ linked $L(C,S)$ times\footnote{For
physical vortex fluxes, which are endowed with a certain thickness
in the directions perpendicular to $S$, the contour $C$ would have to
circumscribe $S$ at distances larger than this thickness in order to
capture the entire vortex flux and not intersect it. As already discussed
further above, in the present treatment, no explicit transverse vortex
thickness will be introduced. Thus, such a qualification does not arise.}
to the surface $S$. Note that the factor $(-1)$ on the right hand side
of eq.~(\ref{lino}) corresponds to the only nontrivial center element
of the $SU(2)$ group, which will be the subject of investigation in the
following. It is in this sense that vortex flux is quantized according
to the center of the gauge group. For higher $SU(N)$ groups, the $(-1)$
on the right hand side of eq.~(\ref{lino}) is replaced by center
elements of those groups. In general, there are several such center
elements and, consequently, several types of quantized vortex flux.

For the purpose of constructing a gauge field
encoding the property (\ref{lino}), it is useful to shift the emphasis
of the description \cite{cont}, namely from the closed surfaces $S$ to 
three-dimensional volumes $\Sigma $ bounded by $S$, i.e. 
$S=\partial \Sigma $. Then (\ref{lino}) can be rewritten as
\begin{equation}
W [C] = (-1)^{I(C,\Sigma )}
\label{isno}
\end{equation}
in terms of the intersection number\footnote{In the 
exponentiated form (\ref{isno}), it is irrelevant whether one counts
intersections without reference to the relative orientation of $\Sigma $
and $C$, or weighted with a sign according to the orientation. Below,
when defining the gauge field itself, it will be necessary to be more
specific, and the orientation will play a crucial role.} $I(C,\Sigma )$ of
$C$ with $\Sigma $.

The random vortex surface model studied here is formulated on a space-time
lattice, on which vortex surfaces are composed of elementary
squares; correspondingly, the volumes $\Sigma $ are composed of
elementary three-dimensional cubes. To specify a volume $\Sigma $, as
yet without orientation, it is sufficient to associate each elementary
three-dimensional cube $K$ on the lattice with a value
$\phi (K) = \pm 1 $, $\phi (K) =1$ signifying that $K$ is not part of 
$\Sigma $, and $\phi (K) = -1$ signifying that it is. The question of how 
a volume $\Sigma $ for a given vortex surface $S$ can be obtained in practice
will be discussed further below; for the time being, let it be assumed
that such a volume has been found (obviously there is some freedom
in choosing $\Sigma $, since it merely must satisfy $\partial \Sigma =S$).
The Wilson loop (\ref{isno}) in this language 
becomes\footnote{Each cube $K$ corresponds to a link on the lattice dual
to the one used in the present construction, that link perpendicularly
piercing the cube $K$ in question. This dual lattice is the lattice one
would use to define standard lattice gauge theory; indeed, if one
transfers the values $\phi (K)$ to the corresponding dual links,
one has precisely constructed a $Z(2)$ lattice gauge configuration
with the vortex content given by $S$. The Wilson loop is then the
product over all links it contains, as is manifest in eq. (\ref{wilu}).
Nevertheless, it should be emphasized that the gauge fields defined in
the following, designed such as to reproduce (\ref{wilu}), contain
considerably more information than the aforementioned $Z(2)$ lattice
configuration. This information, as will become clearer in
section \ref{globsec}, is encoded in the manner in which different cubes
$K_i $ are sewn together at their shared boundaries such as to compose
entire three-dimensional volumes $\Sigma $. These volumes contain
unambiguous global topological information, reflected in the gauge fields
constructed here; by contrast, in the standard lattice formulation, such
topological information is, at best, very implicit. The gauge fields
obtained in the present construction are {\em continuum} gauge fields,
even if they do assume a decidedly cubistic form.}
\begin{equation}
W [C] = \prod_{i: \, K_i \cap C \neq 0} \phi (K_i )
\label{wilu}
\end{equation}
where the product is taken over all cubes\footnote{In the following, when
more than one cube $K$ is being referred to, this will be indicated by
supplying an index label $i$, i.e. $K_i $, where it should be clear from
the context which set the index $i$ runs over.} $K_i $ intersected by $C$.

This must now be translated into the gauge field $A(K)$ associated
with each cube $K$, i.e. a gauge field must be constructed which
satisfies
\begin{equation}
\exp \left( i \int_{P} dx_{\mu } A_{\mu } (K) \right) = \phi (K)
\end{equation}
for a path $P$ in $\mu $-direction intersecting the three-dimensional
cube $K$, assumed here to extend into all directions but the 
$\mu $-direction. Gauge fields $A_{\mu } (K) $ which
reproduce this property can be given as follows. If $K$ is the cube 
extending from the lattice site $\bar{x} $ into all (positive) directions
except the $\mu $-direction, then (with $a$ denoting the lattice spacing
and the third Pauli matrix $\sigma^{3} $ encoding the color structure)
\begin{equation}
A_{\mu } (K) = n(K) \pi \sigma^{3} \delta (x_{\mu } -\bar{x}_{\mu } )
\prod_{\lambda \neq \mu } \theta (x_{\lambda } -\bar{x}_{\lambda } )
\theta (\bar{x}_{\lambda } +a - x_{\lambda } )
\label{akdef}
\end{equation}
i.e. $A(K)$ has support only on $K$ and points into the space-time
direction perpendicular to $K$; furthermore, its strength depends on
an integer $n(K)$ which, at this point, is merely constrained to
satisfy 
\begin{equation}
\exp (i\pi n(K)) = \phi (K) \ ,
\label{ncons}
\end{equation}
i.e. $n(K)$ must be even for $\phi (K) = 1$ and odd for $\phi (K) = -1$.
Apart from this, the integer $n(K)$ still remains to be specified
(which is where the orientation of $\Sigma $ will attain relevance, along
with the necessity of dividing space-time into patches, as already
indicated in section \ref{gensec}). Finally, the color structure of the
gauge field is given by the third Pauli matrix $\sigma^{3} $ in
(\ref{akdef}). As already discussed in section \ref{gensec}, this is a
specification which was already adopted previously
\cite{preptop},\cite{cont} on the level of the vortex field
strengths; it implies that vortex configurations at this stage are cast in
a gauge of the maximally Abelian type. Another strong restriction of the
gauge freedom is of course implied by the space-time form of (\ref{akdef});
the gauge field only has support on a three-dimensional submanifold made
up of elementary three-dimensional cubes in the
lattice\footnote{This general type of gauge field was called 
``ideal vortex field'' in \cite{cont}, with the slight difference that 
in \cite{cont}, the ideal vortex field had support only on $\Sigma $, 
i.e. on the subset of cubes $K_i $ with $\phi (K_i ) = -1$. Below, the
definitions will in fact be brought to match even more closely, by viewing
those $K_i $ which are associated with $\phi (K_i ) = 1$, but which
nevertheless support a nontrivial gauge field (such $K_i $ are unavoidable
if one wishes to exclude Dirac strings, cf. section \ref{globsec}), as two
cubes $K_{i1} , K_{i2} $ with $\phi (K_{i1} ) = \phi (K_{i2} ) = -1$ which
are parallel and slightly displaced from one another; i.e., these are
viewed as additional parts of $\Sigma $ which initially could not be
resolved due to the coarse-grained character of the lattice description.},
as already indicated and discussed in section \ref{gensec}. In fact, apart
from the integer $n(K)$ to be specified in (\ref{akdef}), the only freedom
which remains is the choice of the three-dimensional volume $\Sigma $
spanning the given vortex surface $S$. This freedom corresponds to the
$Z(2)$ part of the $SU(2)$ gauge freedom of $SU(2)$ Yang-Mills
theory \cite{cont}. Again, the question of how $\Sigma $ can be obtained
for a given vortex surface $S$ in practice is deferred for now; suffice
it to state that more than one choice will be explored below.

\subsection{Global topology and space-time patches}
\label{globsec}
In a gauge field formulation of vortex configurations, one needs a
more specific characterization of vortices than is contained in
the property (\ref{lino}); this formally manifests itself in
the a priori freedom in the choice of the integer $n(K)$ in (\ref{akdef}).
Within the random vortex surface model, the only physical degrees of
freedom are elementary vortex fluxes whose gauge field description
satisfies
\begin{equation}
\oint_{C} dx_{\mu } A_{\mu } = \pm \pi \sigma^{3}
\label{avordef}
\end{equation}
for a line integral along an infinitesimal loop $C$ circumscribing the
vortex\footnote{Of course, for physical thick vortices, the relevant
loops would be ones of a size corresponding to the vortex thickness,
such as to capture the entire vortex flux.}.

This property will serve to fix the integer $n(K)$ in (\ref{akdef}) to a
large extent. Eq.~(\ref{avordef}) should be contrasted with the
exponentiated form (\ref{lino}), which would allow any odd integer
multiple on the right hand side\footnote{Additionally, eq. (\ref{lino})
would allow for arbitrary world-surfaces of fluxes corresponding to
even integer multiples of the elementary fluxes (\ref{avordef}).}
of (\ref{avordef}). It should be noted that the specification
(\ref{avordef}) was already used when calculating the topological charge
of vortex configurations \cite{preptop},\cite{cont}; fluxes corresponding
to integer multiples of (\ref{avordef}) of course carry associated
multiple field strengths, which would enter the topological density
$\epsilon_{\mu \nu \kappa \lambda } F_{\mu \nu } F_{\kappa \lambda } $.
Such multiple field strengths were not allowed for in 
\cite{preptop},\cite{cont}; the requirement (\ref{avordef}) thus is not
a new feature of the present investigation. The only freedom in
(\ref{avordef}) is the choice of sign, i.e.~the orientation of the vortex
flux; this can be seen as a vestige of the color rotation freedom present
in the characterization (\ref{lino}) which still remains after fixing the
color direction of the gauge field via the Pauli matrix $\sigma^{3} $
in (\ref{avordef}). In view of the nonorientability of generic vortex
surfaces \cite{bertle}, this residual freedom will be crucial.

This leads to a central issue, namely how the different individual cubes
$K_i $ are sewn together at the elementary squares (plaquettes) they share.
By applying the specification (\ref{avordef}) to a small loop
encircling each plaquette on the lattice, a set of constraints on the
choices of integers $n(K_i )$, cf. eq.~(\ref{akdef}), can be derived.
To be precise, if the plaquette $p$ extends from the lattice site 
$\bar{x} $ into the positive $\mu $ and $\nu $ directions, and
$\lambda ,\kappa $ denote the other two space-time 
directions\footnote{A definite ordering of $\lambda ,\kappa $ is not
necessary for the following construction.}, then
consider the square loop $C(p)$ given by the following sequence of
corners,
\[
C(p) : \ \ \bar{y} \ \ \longrightarrow \ \ 
\bar{y} + a e_{\lambda } \ \ \longrightarrow \ \
\bar{y} + a e_{\lambda } + a e_{\kappa } \ \ \longrightarrow \ \
\bar{y} + a e_{\kappa } \ \ \longrightarrow \ \ \bar{y}
\]
starting and ending at $\bar{y} = \bar{x} + (a/2) (e_{\mu } + e_{\nu } 
- e_{\lambda } -e_{\kappa } )$, where $a$ denotes the lattice spacing
and $e_{\rho } $ the unit vector in the $\rho $-direction. Clearly,
(\ref{avordef}) then implies a constraint on the four cubes $K_i $
whose boundaries contain the plaquette $p$ (and which are thus intersected
by $C(p)$), namely
\begin{eqnarray}
& & \! \! \! \! \! \! \! \!
\oint_{C(p)} dx_{\mu } A_{\mu } = \nonumber \\
& & \ \ \ \ \
= \left[ n ( K_{\lambda } (\bar{x} - a e_{\kappa } ) ) +
n ( K_{\kappa } (\bar{x} ) ) -
n ( K_{\lambda } (\bar{x} ) ) -
n ( K_{\kappa } (\bar{x} - a e_{\lambda } ) ) \right] \pi \sigma^{3}
\nonumber \\
& & \ \ \ \ \ 
\in \ \ \{ 0,\pm \pi \sigma^{3} \}
\label{phicr}
\end{eqnarray}
where $K_{\rho } (\bar{x} )$ denotes the cube extending from $\bar{x} $
into all (positive) directions but the $\rho $ direction. In physical
terms, the plaquette $p$ should carry at most one single unit of
vortex flux (of either orientation). Every plaquette on the lattice
generates one such constraint. It should be noted that this does not
exclude any particular cube $K$ on its own carrying an arbitrary
integer $n(K)$ in eq.~(\ref{akdef}); only from the point of view of 
choosing a gauge in which the gauge fields are as ``smooth'' as possible,
cf.~the general discussion in section \ref{gensec}, it is desirable to keep
the magnitude of $n(K)$ as small as can be achieved; how small this
is in practice will be commented upon below.

\begin{figure}[h]
\centerline{
\hspace{-0.7cm}
\epsfxsize=8.1cm
\epsffile{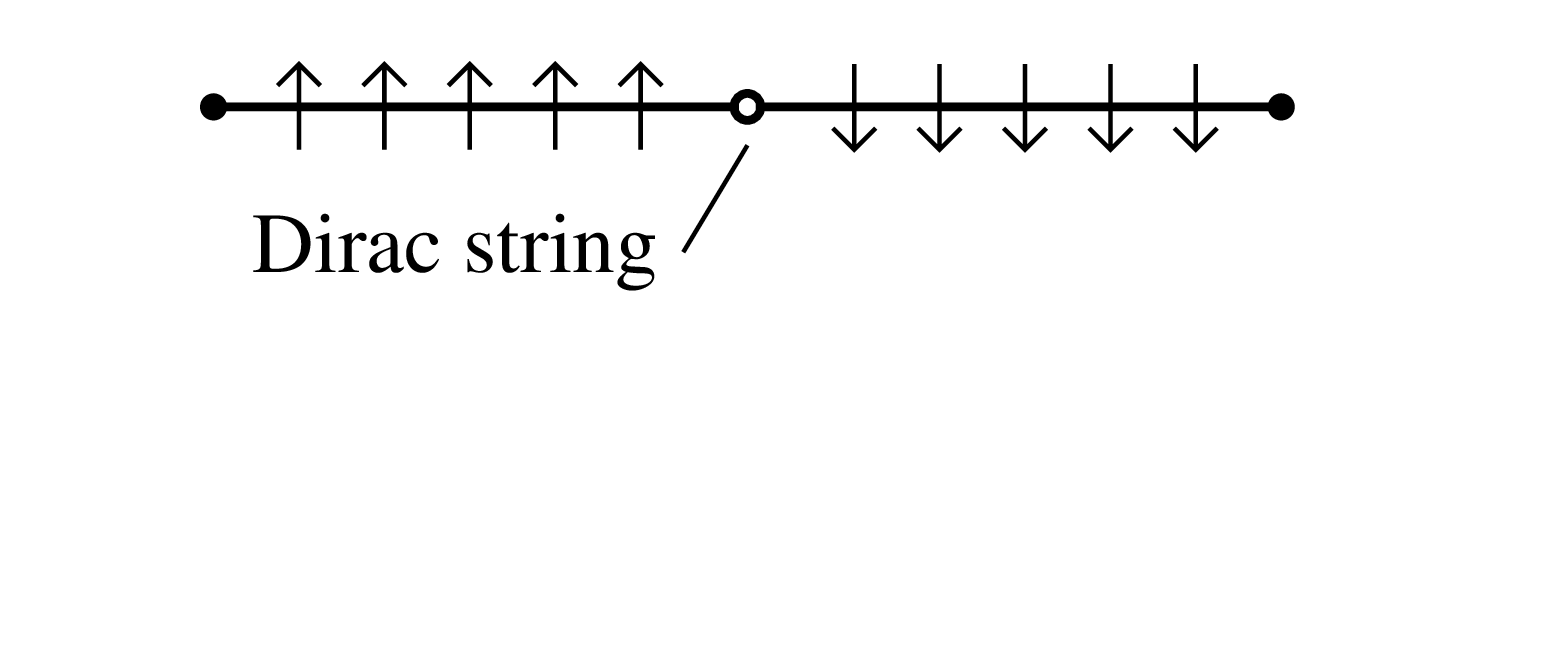}
\hspace{-1.1cm}
\epsfxsize=8.1cm
\epsffile{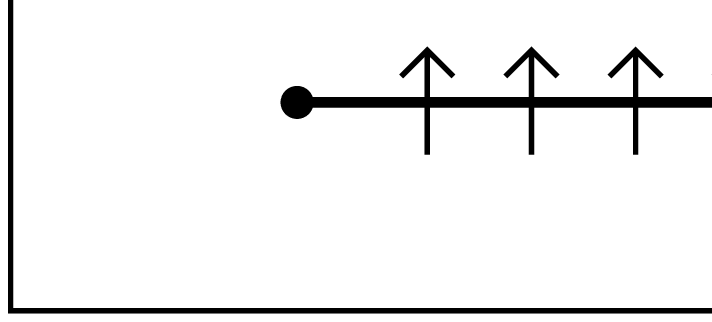}
}
\caption{Nonoriented gauge field support volume $\Sigma $ in a
two-dimensional slice of space-time. In such a slice, vortices show up as
the solid points, volumes $\Sigma $ emanating from them as the solid lines.
The directions of the vortex fluxes at the boundaries of $\Sigma $, fixed
a priori, here happen to be such that the segments of $\Sigma $ emanating
from the different vortices are forced to have mutually incompatible
orientation. This is indicated by the arrows, which
symbolize (directed) line integrals $\int A_{\mu } dx_{\mu } $;
{\em all of the line integrals indicated in the figures yield the same
value}, for definiteness $+\pi \sigma^{3} $. Left: Attempt at a global
definition of the gauge field; the volume $\Sigma $ then necessarily
contains the indicated Dirac string, satisfying the property (\ref{diflu}).
Right: Wu-Yang construction, with space-time subdivided into two patches,
on each of which the volume $\Sigma $ is oriented. Gauge fields on the
two patches are related by the transition function $U$. In the simple
example depicted here, $U$ can be chosen constant in the whole overlap
region shared by the two patches. In generic configurations, the presence
of other vortices nearby forces the space-time region of nontrivial $U$ to
be more localized, namely onto the immediate vicinity of $\Sigma $. This
is discussed in more detail further below.}
\label{fig1}
\end{figure}

With the constraints (\ref{phicr}) relating neighboring cubes $K_i $ on the
lattice, the global topology of the volumes $\Sigma $ made up of such cubes
enters the description. Namely, generic vortex surfaces $S$, and,
concomitantly, volumes $\Sigma $ spanning $S$, are not orientable
\cite{bertle}. Formally, this expresses itself in the property that the
constraints (\ref{phicr}) cannot all be simultaneously satisfied throughout
space-time with a unique choice $n(K_i )$ for all cubes $K_i $ in the
lattice. Rather, for non-orientable $\Sigma $, it is unavoidable to
encounter frustrations in the attempt to satisfy all the constraints,
i.e. one will find some plaquettes $p$ within the volume $\Sigma $ which
yield the value
\begin{equation}
\oint_{C(p)} dx_{\mu } A_{\mu } = \pm 2\pi \sigma^{3} \ ,
\label{diflu}
\end{equation}
thus carrying flux double that of a vortex, cf.~Fig.~\ref{fig1} (left).
Not being intended to appear
as physical flux degrees of freedom within the random vortex surface
model, these are the Dirac string world-surfaces which are unavoidable
in a global, uniquely defined Abelian gauge field describing a
non-oriented vortex surface; their boundaries correspond
to monopole loops on the vortex sheets \cite{cont}. It is clear that
such Dirac strings must arise when the vortex surface $S$ is not
orientable: In that case, there must be lines on the vortex surface
where flux orientation changes (the monopoles), and a globally defined
Abelian gauge field, which satisfies continuity of flux, must therefore
contain Dirac strings supplying the change in vortex flux at the monopoles.

For reasons motivated in section \ref{gensec}, such (Abelian) Dirac strings
are to be excluded from the gauge field in the present framework. Instead,
the Wu-Yang construction will be used: Define the gauge field only locally
on patches, sufficiently small such that all constraints (\ref{phicr})
arising in any particular patch can be simultaneously satisfied.
The different patches are then related via transition functions,
cf.~Fig.~\ref{fig1} (right). The precise construction is as follows. The
patches are the interiors of all the $(2a)^4 $ hypercubes\footnote{I.e.,
the cubes $K_i $ making up the three-dimensional boundary of each hypercube
in question are excluded from the patch.} centered on the sites of the
lattice, $a$ denoting the lattice spacing. This is a natural choice, since
these hypercubes will at the same time be the supports of the quark
basis functions (finite elements) used in the construction of the
Dirac operator matrix further below. In practice, these patches are
sufficiently small to allow all the constraints (\ref{phicr})
(and, of course, (\ref{ncons})) arising on them to be satisfied.
In order to achieve this, it furthermore turns out to be
sufficient to allow for the values\footnote{As already hinted at above,
this minimal choice of $n(K)$ is motivated by the desire to formulate the
gauge field in as ``smooth'' a gauge as possible, cf.~the general remarks
in section \ref{gensec}. On the other hand, it should not come as a
surprise that, in general, one cannot be even more restrictive, i.e.~the
values $n(K)=\pm 2$ should indeed be allowed. Consider two parallel vortex
surfaces {\em of the same orientation} near each other. Then the two
segments of the volume $\Sigma $ emanating from them may in general at
some point (actually, a space-time surface) merge and run superimposed on
one another from this merger surface onwards. Cubes $K_i $ making up this
coincidence volume must be associated with $n(K_i )=\pm 2$; the value
$n(K_i )=0$ by contrast would imply a spurious double vortex (Dirac
string) flux at the merger surface. In principle, the (gauge) freedom
in the space-time choice of $\Sigma $ allows to deform $\Sigma $ such as
to avoid such a coincidence situation; in practice, when deformations are
restricted by the coarse-grained, discrete lattice structure, this may not
be directly possible. However, below, such $n=\pm 2$ volumes will
indeed be reinterpreted as two $n=\pm 1$ volumes running parallel at a
small distance from one another. In effect, this implies that, compared
with its initial construction, the volume $\Sigma $ as a whole may be 
augmented by additional (closed) volumes in order to consistently 
characterize the support of the gauge field $A$.}
$-2 \leq n(K) \leq 2$ in (\ref{akdef}). An actual viable choice of
$n(K_i )$ for each of the $32$ cubes $K_i $ in a given patch is found
by trial and error, starting from an initial cube, assigning values to
neighboring cubes such as to satisfy the corresponding constraints
(\ref{ncons}),(\ref{phicr}), and repeating until integers $n(K_i )$ have
been assigned to all cubes (note that the choice is in general not unique). 
In practice, this procedure uses an insignificant amount of computer
time compared with the subsequent treatment of the Dirac matrix; thus,
it is not necessary to invent more intelligent strategies for finding
a viable set of integers $n(K_i )$.

\subsection{Matching patches - transition functions}
\label{matsec}
\begin{figure}[h]
\centerline{
\epsfxsize=11.1cm
\epsffile{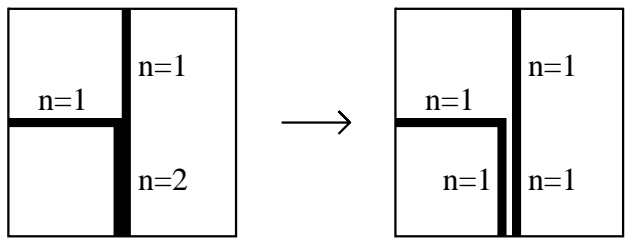}
}
\caption{Splitting of a cube $K$ carrying $n(K)=2$ into two parallel
cubes $K_1 $ and $K_2 $ carrying $n(K_1 )=n(K_2 )=1$.}
\label{fig2}
\end{figure}

To facilitate a simple connection of the different patches via transition
functions, two modifications of the gauge field on each patch are still
necessary. For one, as already hinted at above, any cube $K$ which is
associated with the value $|n(K)|=2$ in at least one of the patches
containing it is replaced by two cubes $K_1 $ and $K_2 $ parallel to $K$,
and displaced from $K$ by a small distance in the direction perpendicular
to the cubes, cf.~Fig.~\ref{fig2}. The new cubes are assigned values
$n(K_1 )$, $n(K_2 )$ in all patches containing $K_1 $ and $K_2 $ as follows:
\begin{eqnarray}
n(K) = 2 & \Longrightarrow & n(K_1 )=1, \ n(K_2 )=1 \nonumber \\
n(K) = -2 & \Longrightarrow & n(K_1 )=-1, \ n(K_2 )=-1 
\label{kreass} \\
n(K) = 0 & \Longrightarrow & n(K_1 )=1, \ n(K_2 )=-1 \nonumber
\end{eqnarray}
Clearly, the new gauge field on the two cubes $K_1 $ and $K_2 $ fulfils
the constraint (\ref{avordef}) defining vortex flux just like the original
gauge field defined on $K$. All that has happened is that the phase picked
up by any line integral intersecting $K$ has been distributed onto the
two cubes $K_1 $ and $K_2 $ which are slightly displaced from $K$. In
effect, the splitting of $K$ corresponds to a deformation of the volume
$\Sigma $ spanning the vortex world-surfaces $S=\partial \Sigma $.
The original $\Sigma $ is augmented by closed volumes; such an operation
does not change $\partial \Sigma $, i.e. the physical content of the
vortex configuration. Rather, it implies a ($Z(2)$) change of the gauge
in which the gauge field is cast. As a result of the assignment 
(\ref{kreass}), only the values $-1 \leq n(K) \leq 1$ are still possible,
at the expense of having introduced additional cubes $K$ into the gauge
field configurations.

\begin{figure}[h]
\centerline{
\epsfxsize=10.1cm
\epsffile{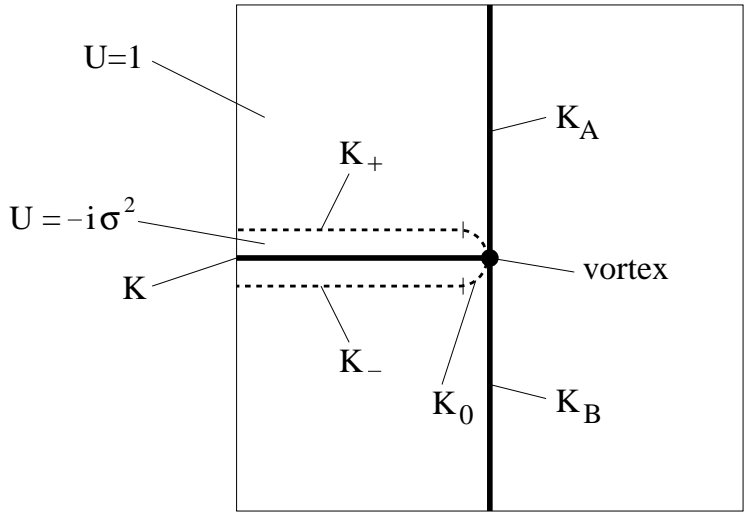}
}
\caption{Space-time patch with three cubes $K$, $K_A $ and $K_B $ which
are part of the gauge field support volume $\Sigma $; in the two-dimensional
slice displayed, these cubes show up as the solid lines. For definiteness,
$n(K_A )=n(K_B )=1$ and $n(K)=-1$; there is a vortex at the center of the
patch. On a neighboring patch which also contains $K$, this cube shall
carry the value $n^{\prime } (K)=1$. The region on which the associated
transition function $U$ is nontrivial is localized as specified in the
figure; this simultaneously implies supplementing the gauge field in the
depicted patch with an auxiliary field $\bar{a} = iU^{\dagger } \partial U$,
see text. The field $\bar{a} $ has support on the volumes $K_{+} $, $K_{-} $
and $K_0 $, which is where $U$ changes as a function of space-time. In
general, it cannot be avoided that $K_0 $ intersects parts of $\Sigma $
(in the absence of $K_A $ and $K_B $, one could still deform $K_0 $ such
as to circumscribe $K$). Such intersections imply spurious fluxes, see text.}
\label{fig3}
\end{figure}

The second change in the gauge field configurations which is still
necessary stems from the following difficulty: In generic vortex
configurations, one does not have the freedom to choose constant
transition functions throughout the overlaps between different patches,
such as in Fig.~\ref{fig1} (right). Instead, the presence of other vortices
nearby forces one to localize the nontrivial parts of the transition
functions to the immediate vicinities of the cubes $K$ being rotated;
in other words, the construction should be local to such an extent that
the transition function can be specified independently for each cube $K$,
cf.~Fig.~\ref{fig3}. However, such more sophisticated transition functions
are space-time dependent and thus call for additional auxiliary gauge
fields in some of the participating patches due to the inhomogeneous
term in the gauge transformation law. Specifically, the modification
consists in adding a pure gauge as follows. For any cube $K$ associated
with the value $n(K)=-1$ on a given patch, consider a space-time region
${\cal E} (K)$ in that patch, bounded by two cubes $K_{+} $ and $K_{-} $
which are parallel to $K$, and which are displaced from $K$ by a small
distance $\epsilon $, as well as a small volume $K_0 $ (of extension
$\epsilon $ in one of its directions) connecting $K_{+} $ and $K_{-} $,
such as depicted in Fig.~\ref{fig3}. Note that any definition of
${\cal E} (K)$ such that ${\cal E} (K)$ contains $K$, but no points more
distant from $K$ than $\epsilon $, serves the purpose of the present
construction. If desired, it is quite admissible to deform $K_0 $ in a way
which excludes from ${\cal E} (K)$ some points which are closer to $K$ than
$\epsilon $. Note also that the distance $\epsilon $ is intended
to be even smaller than the distance between two cubes $K_1 $ and $K_2 $
used in the cube-splitting procedure discussed in the previous paragraph.
Consider furthermore the gauge transformation 
\begin{equation}
U_{ {\cal E} (K)} (x) = \left\{
\begin{array}{lcr}
1 & \mbox{for} & x \notin {\cal E} (K) \\
-i\sigma^{2} & \mbox{for} & x \in {\cal E} (K)
\end{array} \right.
\label{trfel}
\end{equation}
and the gauge field
$\bar{a} (K)=iU_{ {\cal E} (K)}^{\dagger } \partial U_{ {\cal E} (K)} $
induced by $U_{ {\cal E} (K)} $, which has support on the boundary of 
${\cal E} (K)$ described above.

On any patch on which $n(K)=-1$, the field $\bar{a} $ is added to the gauge
field in that patch, i.e.
\begin{equation}
A(K) \longrightarrow \left\{
\begin{array}{lll}
A(K) & \mbox{for} & n(K)=1 \\
A(K)+\bar{a} (K) & \mbox{for} & n(K)= -1
\end{array} \right.
\end{equation}
for each cube $K$ in all patches. This modification does not change any
Wilson loops nor any line integrals (\ref{avordef}) taken in the
configuration; also the physically relevant field strength content of the
gauge field on the patch in question is not changed, apart from the
following subtlety: In general, it is not excluded that $K_0 $ may be forced
to intersect the volume $\Sigma $ which spans the vortex world-surfaces
$S=\partial \Sigma $, and which supports a nontrivial gauge field $A$, 
cf.~Fig.~\ref{fig3}. In this case, a spurious flux is generated on the
intersection surface through the commutator term in the field strength,
$F_{\mu \nu } \propto [A_{\mu } , \bar{a}_{\nu } ]$. Thus, in some cases, 
spurious fluxes of the Dirac string type are reintroduced;
however, the crucial point is that these fluxes have now been
cast in a form in which infrared quark propagation
manifestly is not influenced by them. Namely, consider any coupling
matrix element
\begin{equation}
\int d^4 x \, q_{1}^{\dagger } \, \bar{a} \, q_2
\label{copmix}
\end{equation}
taken between smooth\footnote{The alert reader may notice that, in general,
one will also have to consider the case where one of the quark wave
functions is discontinuous due to the presence of a nontrivial transition
function (\ref{trfel}), e.g. $q_2 =Uq_{2}^{\prime} $ with smooth
$q_{2}^{\prime} $. Here, $q_{2}^{\prime} $ represents a smooth quark
field on an adjacent space-time patch, which must be gauge-transported
to the current patch via the transition function $U$ for the purpose of
evaluating the matrix element. This discussion is deferred to the beginning
of section \ref{qsec}, where the quark basis functions are defined, in
order to be able to treat the issue in a more definite manner. The
statement that integrals of the form (\ref{copmix}) are negligible
will indeed remain true even in this more general case.}
quark wave functions $q_1 $ and $q_2 $. Such an
integral is negligible, since the fields on $K_{+} $ and $K_{-} $ differ
precisely by a sign, and $K_{+} $ and $K_{-} $ are located only a very
small distance $\epsilon $ apart; furthermore, the contribution from
$K_0 $ is of order $\epsilon $ due to the small extension of $K_0 $ in
one of its space-time directions. Thus, the smooth quark modes of an
infrared effective theory are insensitive to the supplementary field
$\bar{a} $. It is not excluded that a more sophisticated construction
exists which avoids the residual spurious field strengths induced by
the additional presence of $\bar{a} $ altogether; however, in view of the
insensitivity of infrared quark propagation with respect to $\bar{a} $, it
does not appear necessary to develop such a construction at this stage. In
a sense, the situation is opposite to the case of the Dirac string fluxes
which one would obtain by naively defining a global Abelian gauge field such
as discussed further above; in that case, the fact that one is working with
a truncated quark basis precludes the solutions of the Dirac equation
adjusting such as to make the Dirac strings invisible. In the present case,
it is precisely the infrared sector of smooth quark wave 
functions which manifestly decouples from any residual spurious
fluxes introduced by the addition of the field $\bar{a} $.
Note, however, that this is one of the main points
preventing a straightforward generalization of the present gauge field
definition to explicitly thickened vortex fluxes and gauge field volumes.
Such a generalization would entail finite $\epsilon $, and as a
consequence, the spurious additional field strengths induced by adding
the field $\bar{a} $ as discussed here could no longer be argued to decouple.
Thus, a more sophisticated construction of the gauge fields on each patch
and the associated transition functions would need to be specified.

Note furthermore that, at this point, one still has a certain amount
of freedom as regards the global definition of the gauge field, which
will be discussed in more detail below in section \ref{optsec}.
In particular, it is still possible to vary the space-time density of
nontrivial matchings between neighboring
patches, i.e.~nontrivial transition functions. Comparing the results
obtained with constructions which mutually differ in this respect will
give an indication of the influence of the rough disorder in the gauge
field associated with the presence of those nontrivial transition functions.

Having prepared the gauge field configurations on each space-time patch
as described above, it is straightforward to give
the corresponding transition functions between patches. Consider two
patches, distinguished in the following by denoting all quantities in
one patch with a prime and their counterparts in the other patch without
a prime. Consider furthermore the space-time region in which the two 
patches overlap and the set of cubes $K_i $ located there, associated
with values $n(K_i )$ and $n^{\prime } (K_i )$ in the two patches. The
associated transition function in that overlap region then is
\begin{equation}
U=\prod_{K_i } U^{[n(K_i )-n^{\prime } (K_i )]/2}_{ {\cal E} (K_i ) }
\label{uputo}
\end{equation}
with $U_{ {\cal E} (K)} $ given by (\ref{trfel}), i.e. the gauge fields
$A$ (including, if present, the auxiliary field $\bar{a} $) and the quark
fields $q$ in the two patches are related by
\begin{eqnarray}
A^{\prime } &=& UAU^{\dagger } -iU\partial U^{\dagger } \\
q^{\prime } &=& Uq
\label{quatra}
\end{eqnarray}

\subsection{Options in the gauge field construction}
\label{optsec}
The gauge field construction presented above still allows for different
model options in two respects. In the numerical work further below,
a variety of these options will be explored in order to be able to
appreciate the dependence of the results on the gauge field modeling,
which, after all, contains a range of ambiguities and idealizations
as presented in the preceding sections.

One of the options one still has lies in the choice of the volume
$\Sigma $ spanning the vortex world-surfaces. A particular such volume
$\Sigma $, the construction of which was repeatedly deferred above,
can be obtained as follows. As discussed in detail in \cite{selprep},
Monte Carlo updates of vortex configurations in the present model are
always carried out simultaneously on all six plaquettes making up the
surface of an elementary three-dimensional cube in the lattice; this
serves to keep the vortex surfaces closed as they are being updated.
As a result, a volume $\Sigma $ whose boundary defines the vortex surfaces
can be trivially obtained by keeping track of the cubes being updated.
To be precise, if one starts with an empty lattice, $\phi (K_i ) =1$
for all three-dimensional cubes $K_i $ in the lattice; whenever an update
of (the surface of) a cube $K$ is subsequently accepted, one concomitantly
changes $\phi (K) \rightarrow -\phi (K)$. In this way, at any point in
the calculation, a viable $\Sigma $, made up of the set of $K_i $ with
$\phi (K_i ) =-1$, is known.

On the other hand, this is not the only possible $\Sigma $. Instead of
working directly with the above $\Sigma $, which will generally be rather
random and rough, it is possible to first apply a smoothing algorithm to 
$\Sigma $. Since the only physically relevant information lies in the
boundary of $\Sigma $, it is admissible to add to $\Sigma $, or take away
from $\Sigma $, arbitrary closed three-volumes. In practice, this implies
sweeping through the lattice, considering each elementary four-cube
in turn, and simultaneously updating all eight three-cubes $K_i $
making up its boundary, i.e. $\phi (K_i ) \rightarrow -\phi (K_i )$,
whenever this leads to a reduction in the total volume of $\Sigma $.
The two options for $\Sigma $ are related by a ($Z(2)$) gauge
transformation which however does not vary smoothly in space-time.
Consequently, invariance of the Dirac spectrum is not guaranteed in a
calculation using a truncated quark basis, as already mentioned in
section \ref{gensec}. Thus, the two options represent different possible
models which will be explored in the numerical work below; the options will
be referred to as ``random $\Sigma $'' and ``smooth $\Sigma $'',
respectively. Of course, under the aspect of formulating the gauge
field in as smooth a gauge as possible in order to minimize artefacts
stemming from rapidly varying gauge transformations, the smooth option
seems the preferred one.

The second option one still has in the construction of the gauge field
is the following: Above, the gauge field $A$ is initially constructed
independently on each space-time patch. As a result, the relative
color orientation of adjacent patches, ultimately encoded in the
transition functions on the respective overlap regions, is random;
there is a high density of overlap regions supporting nontrivial
transition functions. However, for any given patch,
the transformation $A\rightarrow -A$ leads to an equally admissible
gauge field on that patch; this simply corresponds to a gauge
transformation $U=-i\sigma^{2} $ constant throughout the patch.
Thus, after the initial construction of the gauge field $A$ (but before
the addition of the auxiliary gauge field $\bar{a} $ discussed in
section \ref{matsec}), one has the alternative of performing sweeps through
the lattice in which one considers a transformation of the gauge field
$A\rightarrow -A$ on each individual patch in turn. Acceptance of such a
transformation can be biased such that as many as possible of the
resulting transition functions to neighboring patches become trivial. In
other words, the transition functions are made maximally smooth. In the
language of globally defined gauge fields with monopole and Dirac string
singularities, this corresponds to minimizing the space-time
density of such singularities. Accordingly, the option of fixing
the gauge field orientation at random on the patches will below be
referred to as the ``random monopole'' gauge, whereas the option of
maximally smooth transition functions will be referred to as the
``minimal monopole'' gauge. Note that such an alternative, either choosing
the monopoles at random or minimizing their density, was already
considered in \cite{preptop},\cite{bertop},\cite{berproc}. The topological
susceptibility turned out to be virtually independent of the monopole
density, a fact which could be understood in terms of the space-time
properties of generic vortex surfaces
\cite{preptop},\cite{bertop},\cite{berproc}.

\section{Vortex-Quark coupling}
\subsection{Quark basis and Dirac matrix}
\label{qsec}
As already indicated further above, the Dirac equation will be solved
using the finite element method. This means that the truncated quark
basis is constructed using space-time wave functions $f(x-x_0 )$
associated with, and localized around, each lattice site $x_0 $
(the Dirac and color structure will be supplemented further below).
The wave function $f$ is a product over the four space-time directions,
\begin{equation}
f(z) = \prod_{i=1}^{4} h(z_i )
\label{baspro}
\end{equation}
and $h$ is piecewise linear,
\begin{equation}
h(t) = \left\{
\begin{array}{lll}
t+a & \mbox{for} & -a\leq t\leq 0 \\
a-t & \mbox{for} & 0\leq t\leq a \\
0 & \mbox{else} &
\end{array}
\right.
\end{equation}
where $a$ denotes the lattice spacing.
To be precise, the space-time wave function associated with the lattice
site $x_0 $ takes the form $f(x-x_0 )$ {\em on the patch which is
simultaneously associated with $x_0 $}. Viewed from other patches, this
form is of course modified by the corresponding transition
functions\footnote{Note that this means that quark basis functions are
only strictly smooth on the patch they are defined on, whereas, viewed
from the color frame associated with a neighboring patch, they in general
exhibit non-smooth behavior near selected three-dimensional cubes $K$;
after all, the corresponding transition functions can vary rapidly on the
very small length scale $\epsilon $ entering the definition of
${\cal E} (K)$ in (\ref{trfel}). As one moves on such a cube $K$ from
one lattice site to the next, a general quark wave function, linearly
combined from the basis functions, continuously interpolates between a
form which is strictly smooth on the patch associated with the initial
site to a form which is strictly smooth on the patch associated with the
final site. It should be emphasized that even with this generalized (not
strictly smooth) ansatz for the quark wave functions, the statement made
in connection with eq.~(\ref{copmix}) remains true, i.e.~the auxiliary
field $\bar{a} $ cancels in all Dirac matrix elements and in this sense
does not influence quark propagation. This follows from
the property that the part of the wave function rapidly varying around
a given cube $K$, viewed as a function of the coordinate perpendicular
to $K$, is symmetric around $K$. As a consistency check, if one considers
the free Dirac equation in patched space-time with nontrivial transition
functions and the corresponding auxiliary fields $\bar{a} $, the Dirac
matrix is manifestly identical to the one with trivial transition functions.
In the framework of the Wu-Yang construction, the transition functions are
nothing but a part of the choice of gauge; the fact that they (and, as a
consequence, the quark wave functions) are not strictly smooth is merely a
specific instance of the model compromise already discussed in
section \ref{gensec}, namely that the fields in the present construction
are not given in the smoothest possible gauge, but that the gauge is partly
dictated by considerations of manageability.}. An exception to the above
form occurs in the case when $x_0 $ is located on the boundary where the
space-time torus is sewn together, i.e. when at least one of the vector
components of $x_0 $, say $x_{0j} $, vanishes. In this case, the
corresponding factor(s) $h(z_j )$ in (\ref{baspro}) is (are) replaced by
\begin{equation}
\tilde{h} (z_j ) = \left\{
\begin{array}{lll}
-a-z_j & \mbox{for} & -a\leq z_j \leq 0 \\
a-z_j & \mbox{for} & 0\leq z_j \leq a \\
0 & \mbox{else} &
\end{array}
\right.
\end{equation}
In other words, the quark wave functions are forced to obey antiperiodic
boundary conditions in all directions. Physically, this is only really
necessary in the Euclidean time direction; the space directions on the
other hand should always be taken sufficiently large such that bulk
physics does not depend on the behavior at the boundaries. Note that,
as far as the torus periodicity is concerned, the gauge fields discussed
further above obey periodic boundary conditions; translation along one
of the torus directions by its length leads back to the same patch,
with the same gauge field.

A further comment is in order concerning space-times of extent $a$ in 
the Euclidean time direction.
While the vortex surfaces and the associated gauge fields can be
straightforwardly defined on such a space-time, the truncated quark
basis should include the two lowest Matsubara frequencies, which are
degenerate, on an equal footing. Thus, in practice, while the vortex
configurations are still generated on the original lattice with
spacing $a$, these configurations are subsequently copied onto a
lattice with two spacings $a/2$ in the time direction in order to
allow for an (approximate) accomodation of both of the basis functions
corresponding to the lowest Matsubara frequencies.

The above set of wave functions $f$ is supplied for each color and Dirac
component, i.e. the complete quark basis can be labeled as
\begin{equation}
q(b,i,x_0 ) = \vec{c}_{b} \vec{d}_{i} f(x-x_0 )
\end{equation}
where $\vec{c}_{b} $ is the $b$th unit vector in (two-dimensional)
color space, $\vec{d}_{i} $ is the $i$th unit vector in (four-dimensional)
Dirac spinor space, and $x_0 $ is a site on the space-time lattice.
Note that quark wave functions are linearly combined from these
basis vectors using, in general, {\em complex} coefficients.
In this basis, it is straightforward to calculate analytically the
matrix elements of the (Euclidean) Dirac operator
\begin{equation}
\hat{D} \! \! \! \! / \, =
\gamma_{\mu } (i\partial_{\mu } -A_{\mu } )
\end{equation}
using the gauge field $A$ constructed in section \ref{gfsec}. This becomes
particularly simple when the diverse small lengths introduced in
section \ref{matsec} are in fact taken to be negligibly small, specifically,
the distance separating the two parallel cubes $K_1 $ and $K_2 $ in
eq.~(\ref{kreass}), and the (even smaller) width $\epsilon $ of the domain
${\cal E} (K)$ used in eq.~(\ref{trfel}). The Dirac matrices were in
practice taken in the chiral representation,
\begin{equation}
\gamma_{0} = \left( \begin{array}{cc} 0&1\\ 1&0 \end{array} \right)
\ \ \ \ \ \ \ \ 
\gamma_{i} = \left( \begin{array}{cc} 0&-i\sigma^{i} \\
i\sigma^{i} &0 \end{array} \right) \ .
\end{equation}
When evaluating matrix elements between states defined on adjacent
patches, of course the corresponding transition function must be used
to transport all functions involved to a definite color frame, i.e.
for unequal $x_0 ,x_{0}^{\prime } $,
\begin{equation}
\langle b^{\prime } i^{\prime } x_{0}^{\prime } |
\hat{D} \! \! \! \! / \, | b i x_0 \rangle =
\vec{c}_{b^{\prime } } \vec{d}_{i^{\prime } } \int d^4 x\,
f(x-x_{0}^{\prime } ) U(x;x_{0}^{\prime } ,x_0 )
\hat{D} \! \! \! \! / \, f(x-x_0 ) \vec{c}_{b} \vec{d}_{i}
\end{equation}
where $U(x;x_{0}^{\prime } ,x_0 )$ denotes the transition function
connecting the patches centered at $x_0 $ and $x_{0}^{\prime } $,
the assignment of primed and unprimed patches having been chosen
as in eqs.~(\ref{uputo})-(\ref{quatra}).

Note that the quark basis used here is not orthogonal, i.e. the
overlap matrix $M$ made up of the matrix elements
\begin{equation}
\langle b^{\prime } i^{\prime } x_{0}^{\prime } | b i x_0 \rangle
\end{equation}
is not the unit matrix. Thus, to obtain the eigenvalues $\lambda_{n} $ of
the Dirac operator, one must solve the matrix equation
\begin{equation}
D \! \! \! \! / \, z_n = \lambda_{n} M z_n
\end{equation}
Furthermore, since the square of the Dirac operator and the overlap
matrix $M$ are block-diagonal in the Dirac indices, in practice it is
advantageous to instead diagonalize the squared Dirac operator, 
i.e.~to consider the equation
\begin{equation}
M^{-1} D^{\dagger } M^{-1} D z_n =  \lambda_{n}^{2} z_n
\label{dirbl}
\end{equation}
which is only two-dimensional in Dirac spinor space, $D$ denoting the
matrix of the operator
\begin{equation}
\hat{D} = i\partial_{0} -A_0 -i\sigma^{i} (i\partial_{i} -A_i )
\end{equation}
Note that the Dirac matrix constructed here is manifestly Hermitean
and chirally symmetric, i.e. its eigenvalues are real, and when they
are nonzero, they trivially appear in pairs of opposite sign.
On the other hand, since the quark basis is finite-dimensional, no
{\em exact} zero-mode solutions occur. As a consequence, the spectrum
obtained here by construction strictly only contains the non-zero mode sector
of vanishing chirality. This is different from lattice Dirac operators,
in which both chiral symmetry is broken and the quark basis is
truncated; the continuum limit then corresponds to a {\em combined}
restoration of chiral symmetry and basis completeness - this in principle
also allows one to capture exact zero modes in the 
limit\footnote{Ultimately, this is connected with the fact that the
standard lattice cutoff regularizes the theory {\em gauge-invariantly} in
the ultraviolet, as opposed to the fixed momentum cutoff implied by an
infrared effective framework such as discussed here.}. By contrast, in
the present continuum approach, the limit of exact chiral symmetry
is taken already at fixed basis truncation. With this order of limits,
chiral modes cannot be recovered and must be obtained separately by
different means. Such an investigation, e.g. via the spectral flow of
the Dirac operator in the presence of an explicit chiral symmetry-breaking
parameter, lies outside the scope of the present work. For the purpose
of describing the chiral condensate, the exact zero mode sector is
irrelevant in the infinite volume limit, simply because the topological
susceptibility $\langle Q^2 \rangle /V$ is a finite quantity ($Q$ denotes
the topological charge, each unit of which is expected to be associated
with a quark zero mode via the Atiyah-Singer index theorem, and $V$ is
the space-time volume under consideration). As a consequence, the
contribution of the exact zero modes to the total Dirac spectral density
becomes negligible as $V\rightarrow \infty $.

\subsection{Remarks on topology and torus twist}
Before applying the construction presented in the previous sections to
the configurations of the random vortex surface model, a comment on
their topology is in order. The topological charge of these configurations
is quantized in half-integer units \cite{preptop}. Usually, such
half-integer charge is thought to require twisted boundary conditions for
the gauge fields on the torus \cite{baal},\cite{gonzar1},\cite{baal2}.
These are boundary conditions of the type
\begin{equation}
A(x+L_i ) = A^{U_i (x)} (x)
\label{totwi}
\end{equation}
where $L_i $ is a vector in the $i$-direction of the length of the torus
in that direction, and $U_i $ denotes a gauge transformation. The latter
must obey the consistency condition
\begin{equation}
U_i (x) U_j (x+L_i ) = \pm U_j (x) U_i (x+L_j )
\label{twibc}
\end{equation}
in order to give a unique, well-defined relation between $A(x)$ and
$A(x+L_i +L_j )$. The case of a minus sign on the right hand side
of (\ref{twibc}) is referred to as a twisted boundary
condition\footnote{In the vortex language, the mechanism by which twisted
boundary conditions allow for half-integer topological charge takes the
following form. Twisted boundary conditions permit the existence of a
single vortex world-surface corresponding to a two-dimensional plane
in the torus. On its own, such a plane does not represent the boundary
of any three-dimensional volume, and thus cannot occur for standard
periodic boundary conditions. In the presence of two such planes which
are completely perpendicular to each other, there is precisely one
intersection point of these planes on the torus, which represents the
lone contribution to the topological charge \cite{preptop},\cite{cont}, of
modulus $1/2$. Note that the vortex surface configuration in this case is
oriented; by contrast, for compact vortex surfaces which do not rely
on specific space-time boundary conditions, nonvanishing topological
charge requires non-orientedness \cite{cont}.}. In particular,
fundamental quark fields cannot be defined on a torus with such
twisted boundary conditions; applying the gauge transformations
on either side of (\ref{twibc}) to a quark field $q(x)$ implies
\begin{equation}
q(x+L_i +L_j ) = -q(x+L_i +L_j ) \ ,
\label{inconsq}
\end{equation}
i.e.~$q=0$. 

In contrast to this, the construction presented in the previous
sections uses purely periodic boundary conditions for the gauge fields
on the torus, and in particular also allows the definition of (antiperiodic)
quark fields. With the choice (\ref{uputo}) of transition functions, the
quark fields are uniquely defined on every patch and no inconsistency
such as (\ref{inconsq}) can arise. The construction includes arbitrary
vortex world-surface configurations generated in the random vortex
surface model, in particular ones with half-integer topological charge.

This does not contradict the usual statement that half-integer topological
charge on a torus requires twisted boundary conditions, because the
space-time manifold used in the construction {\em is not a torus in the
strict sense.} Rather, the spurious flux surfaces induced by the
auxiliary fields $\bar{a} $ discussed in section \ref{matsec}, which in 
general cannot be avoided, must be excised from space-time, leading to a
complicated, multiply connected structure\footnote{Note also that, in such
a multiply connected, patched space-time, it is a priori not clear that
the connection between topological charge and zero modes of the Dirac
operator, i.e.~the index theorem, takes the same simple form as in a
standard one-patch space-time.}. By contrast, the case usually
referred to as a twisted torus corresponds to the specific case where
there are no such excisions, and there is only a single patch encompassing
all of space-time, merely with
nontrivial transition functions, cf.~eq.~(\ref{totwi}), where opposite
ends of the patch are sewn together to make up the toroidal space-time
structure. This represents a very rigid and direct relation between the
global shape of space-time and the patching, and the combination of the
two leads to the constraint (\ref{twibc}). On the other hand, in the present
more flexible construction, the patching is completely divorced from the
global torus structure; already the {\em local} structure of space-time,
on scales related to the strong interaction scale, is topologically
nontrivial in a dynamically determined fashion. In this setting, the
additional toroidal periodicity constraints are not as crucial in
determining the global topological properties of the vortex configurations
as in the one-patch, twisted torus case (nevertheless, these periodicity
constraints are still there, with the associated finite size effects).

\begin{figure}[h]
\centerline{
\epsfysize=6.5cm
\epsffile{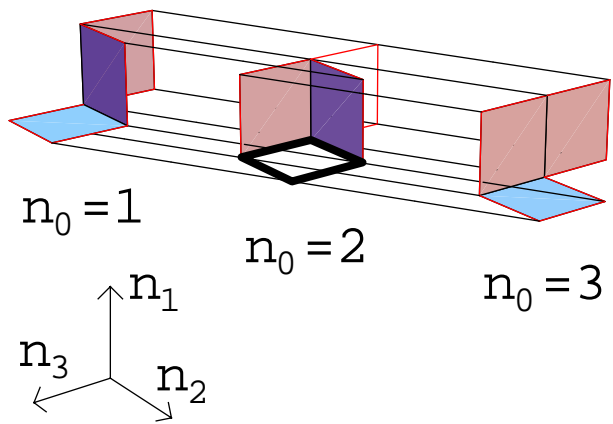}
}
\caption{Sample closed vortex surface configuration. At each lattice time
$n_0 $, shaded plaquettes are part of the vortex surface. These
plaquettes are furthermore connected to plaquettes running in time
direction; their location can be inferred most easily by keeping in
mind that each link of the configuration is connected to exactly
two plaquettes (i.e. the surface is closed and contains no intersection
lines). Note that the two non-shaded plaquettes at $n_0 =2$ are {\em not}
part of the vortex; only the sets of links bounding them are.
These are slices at $n_0 =2$ of surface segments running in time
direction from $n_0 =1$ through to $n_0 =3$. Sliced at $n_0 =2$, these
surface segments show up as lines. The surface is non-orientable; the
thick line at $n_0 =2$ indicates a minimal non-contractible (monopole) loop
on the surface at which the orientation is inverted. Note also that the
topological charge $Q=1/2$ of the configuration is carried purely by
surface writhings; there are no surface self-intersection points. The
author gratefully acknowledges R.~Bertle and M.~Faber for providing the
MATHEMATICA routine with which the image was generated.}
\label{fighalb}
\end{figure}

A particular example of a vortex configuration illustrating this point
is given in Fig.~\ref{fighalb}. This configuration has topological
charge $Q=1/2$ (for details on evaluating the topological charge of a vortex
configuration, cf.~\cite{preptop}). On the other hand, this world-surface
can be represented as the boundary of a three-dimensional volume which
is localized in the region of the vortex. Thus, within the present
construction, this configuration can be encoded in a gauge field $A$ with
support in a compact domain, and $A=0$ in all directions if one moves away
from this domain. Therefore, this configuration can be defined on
space-times of any shape, without relying on special assumptions concerning
the behavior of the gauge fields at the space-time boundaries. However, the
surface is non-orientable; a possible position of the associated
(non-contractible) monopole loop on the vortex world-surface is given by
the thick line in Fig.~\ref{fighalb}. This implies that, within the present
construction, a surface spanning the monopole loop, carrying spurious
nonvanishing field strength, must be excised from space-time. Concomitantly,
the configuration cannot be defined on a single space-time patch; it
requires at least two space-time patches, related by nontrivial transition
functions. In this way, the configuration induces a nontrivial space-time
topology within the region occupied by it; the ``twist'' permitting
topological charge $1/2$ is localized in that region, instead of being
encoded in nontrivial (twisted) torus boundary conditions.

At this point, there seems to be no apparent reason to exclude the
possibility of an alternate gauge field description of such configurations
which indeed transfers the twist required for half-integer topological
charge into twisted torus boundary conditions. Note e.g.~that the
nonabelian theory allows for pure gauges which, in their diagonal color
components, contain Dirac monopole loops spanning open Dirac string
world-surfaces (by contrast, in the Abelian gauge theory, only closed
Dirac strings are pure gauges). An explicit example of such a pure
gauge is given in \cite{cont}; it contains long-range nonabelian gauge
fields. Thus, one could envisage using such gauges to cancel any spurious
Dirac string type fluxes in a gauge field description of the configuration
depicted in Fig.~\ref{fighalb}, at the expense of introducing nonvanishing
gauge fields at large distances; these in general will be influenced by the
space-time boundary conditions.

Note furthermore that even the fact that
a twisted torus does not directly permit a definition of fundamental
quark fields should be seen more as a practical disadvantage than a
fundamental one. By piecing copies of a twisted torus together such as
to make up a larger torus with appropriate integer multiple extensions
in the four space-time directions, one arrives at a description in terms
of gauge fields with periodic boundary conditions on the larger torus.
On this torus, quark fields can then be straightforwardly defined
\cite{gonzar1},\cite{gonzar2}. The price one pays consists in having to
work on a larger space-time manifold; of course, from the practical point
of view of numerical Monte-Carlo calculations, this can be a decisive
disadvantage.

\section{Quenched Dirac spectrum of the random vortex surface model}
\label{ressec}
In the following, results for the quenched ensemble average of the Dirac
spectrum in the framework of the random vortex model of infrared
Yang-Mills theory will be presented. The dynamics determining this
model ensemble and its physical interpretation are discussed at length
in \cite{selprep}. The model describes closed vortex world-surfaces
composed of elementary squares (plaquettes) on a hypercubic lattice.
An ensemble of such world-surfaces is generated using Monte Carlo methods,
where, in order to keep the surfaces closed, an elementary update of a
vortex world-surface configuration always affects all six plaquettes making
up the surface of a three-dimensional cube in the four-dimensional lattice;
any of those six plaquettes which were not part of a vortex surface before
the update become part of a vortex surface after the update, and vice versa.
As mentioned in section \ref{optsec}, keeping track of the updated
three-cubes directly allows the construction of the three-volume $\Sigma $
which is needed as input for the definition of a corresponding vortex gauge
field, cf.~section \ref{locsec}. The acceptance criterion, i.e.~the
Boltzmann factor $\exp (-S)$, for any given update depends on the curvature
of the vortex surfaces. An additive action increment $c=0.24$ is contributed
to $S$ by each pair of vortex plaquettes on the lattice sharing a link,
but not lying in the same plane; the value of $c$ was chosen such as
to quantitatively reproduce the confinement properties of $SU(2)$
Yang-Mills theory \cite{selprep}. This vortex ensemble simultaneously
yields a result for the topological susceptibility which quantitatively
agrees with full $SU(2)$ Yang-Mills lattice calculations \cite{preptop}.

\begin{figure}[h]
\centerline{
\epsfysize=9cm
\epsffile{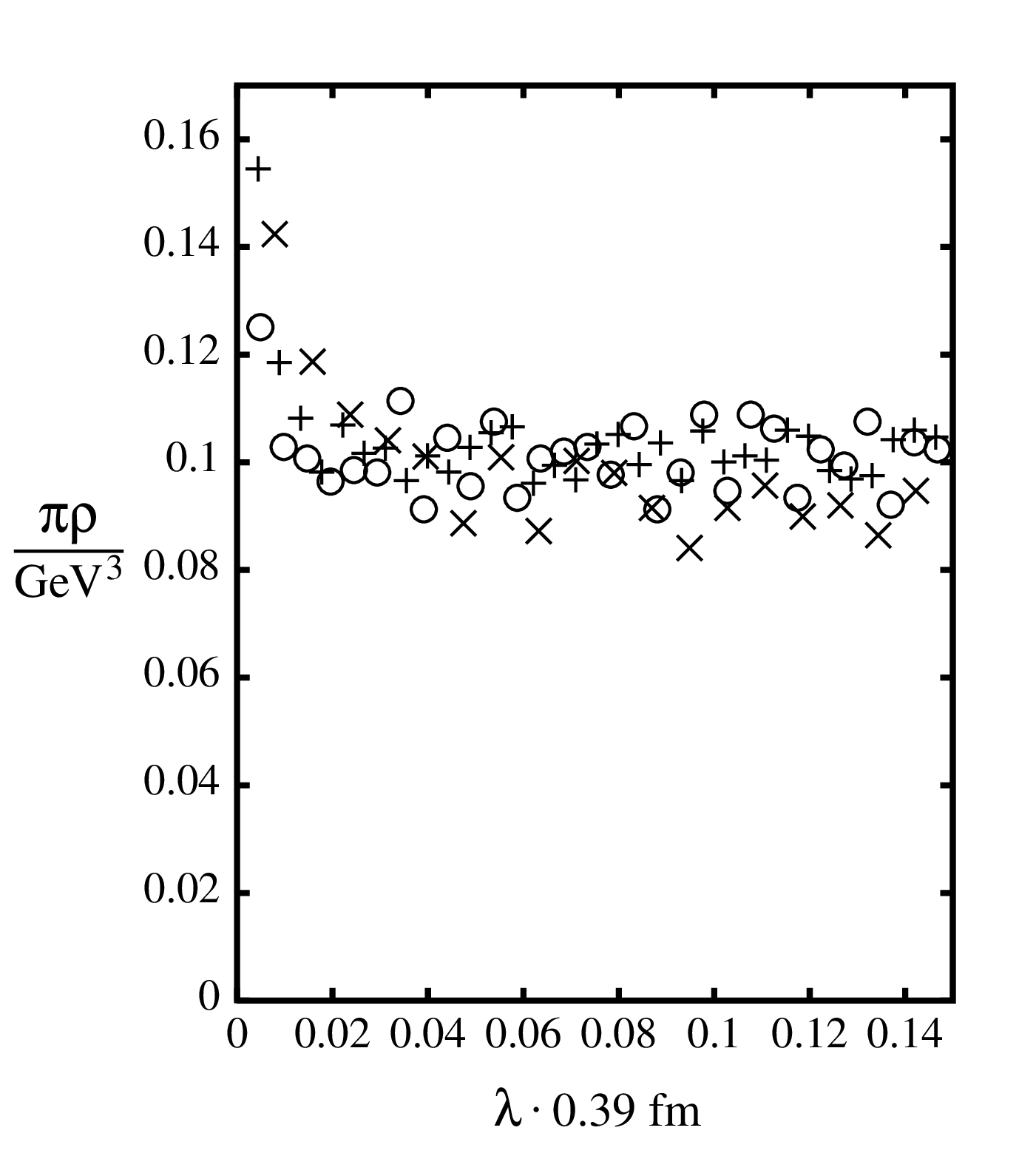}
\hspace{-0.3cm}
\epsfysize=9cm
\epsffile{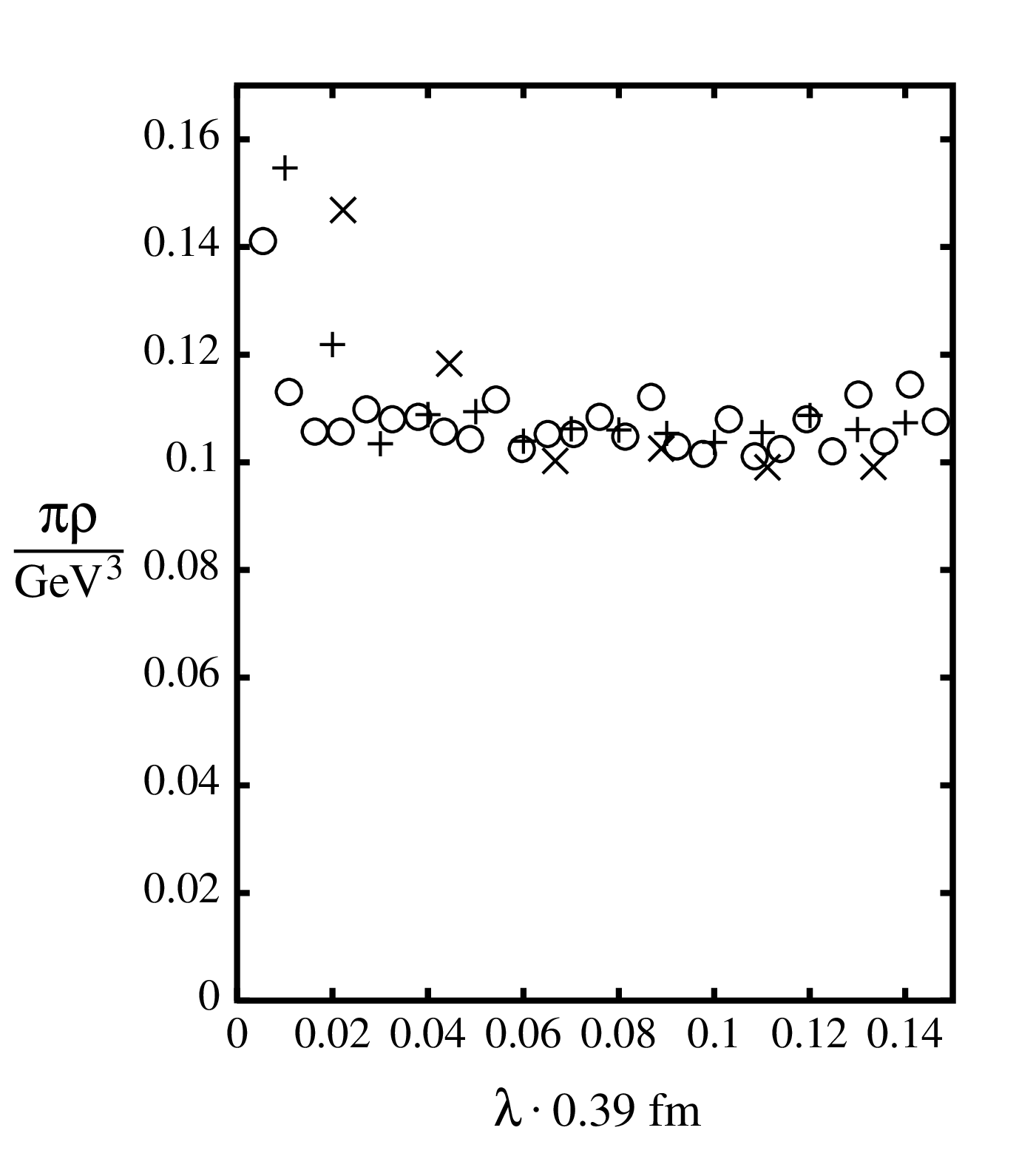}
}
\caption{Dependence of Dirac spectral density $\rho $ on the extension of
space-time. Left: Gauge field constructed using random monopoles and
random $\Sigma $; data symbolized by $\times $ corresponds to a $3^4 $
lattice, crosses correspond to a $4^4 $ lattice, and circles to a $5^4 $
lattice. The reader is reminded that the lattice spacing is $0.39 \, $fm.
Right: Gauge field constructed using minimal monopoles and random
$\Sigma $; data symbolized by $\times $ corresponds to a $3^3 \times 2$
lattice, crosses correspond to a $4^3 \times 2$ lattice, and circles to a
$6^3 \times 2$ lattice. Note that an extension of the lattice in Euclidean
time direction of two spacings is equivalent to a temperature of
$T=0.83\, T_c $.}
\label{sizfig1}
\end{figure}

\begin{figure}[h]
\centerline{
\epsfysize=9cm
\epsffile{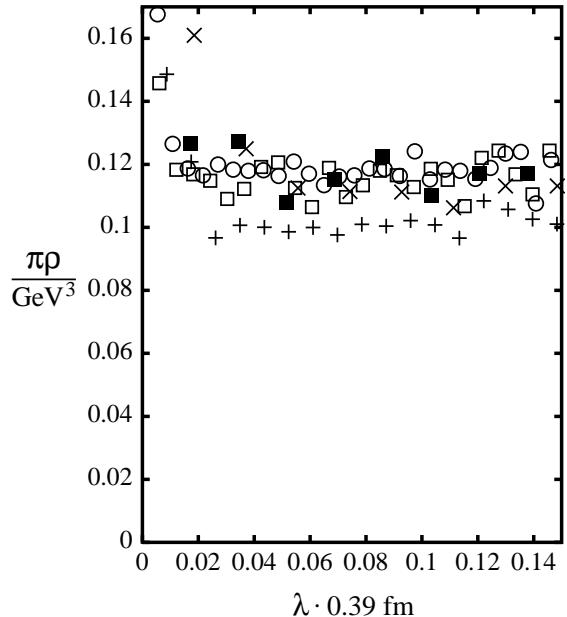}
}
\caption{Dependence of Dirac spectral density $\rho $ on the extension of
space-time. The gauge field is constructed using minimal monopoles and
smooth $\Sigma $. Data symbolized by $\times $ corresponds to a
$3^3 \times 2$ lattice, crosses correspond to a $4^3 \times 2$ lattice,
circles to a $5^3 \times 2$ lattice, open boxes to a $6^3 \times 2$ lattice,
and filled boxes to a $7^3 \times 2$ lattice.}
\label{sizfig2}
\end{figure}

Given a particular vortex configuration, via the associated three-volume
$\Sigma $, the Dirac matrix (in practice, a $2\times 2$ Dirac block of
its square, cf.~eq.~(\ref{dirbl})) can be constructed as described in
sections \ref{gfsec} and \ref{qsec}. The 30 lowest-lying eigenvalues of this
matrix were obtained using the ARPACK package \cite{arpack}, and (their
positive square roots) binned according to their magnitude, yielding the
density of eigenvalues (per unit space-time volume) $\rho (\lambda ) $ of
the Dirac operator\footnote{Strictly speaking, this yields the part of the
spectral distribution located on the positive eigenvalue axis. However, due
to manifest chiral symmetry, the distribution as a whole is symmetric about
eigenvalue zero; therefore, the distribution on the negative eigenvalue
axis does not need to be considered explicitly.}. The behavior of this
distribution near $\lambda =0$ is related to the chiral condensate
$\langle \bar{q} q\rangle $ via the Casher-Banks formula
\cite{cashb},\cite{leuts}
\begin{equation}
\langle \bar{q} q\rangle = 
-\lim_{V\rightarrow \infty }
2m\int_{0}^{\infty } d\lambda \, \frac{\rho (\lambda )}{m^2 +\lambda^{2} }
\stackrel{m\rightarrow 0}{\longrightarrow }
-\pi \rho (0)
\label{cabaf}
\end{equation}
where $m$ denotes the quark mass and $V$ the space-time volume. A range of
results obtained in this way is displayed in Figs.~\ref{sizfig1} and
\ref{sizfig2}, with emphasis on the variation with volume. The two main
features of these eigenvalue distributions are a bulk behavior which
extrapolates to a finite non-zero value at $\lambda =0$, and an additional
enhancement for very small eigenvalues which presumably signals the onset of
divergent behavior at $\lambda =0$. In view of (\ref{cabaf}), such a
divergence implies a diverging chiral condensate in the chiral limit
$m\rightarrow 0$; at finite quark mass $m$, the enhancement near
$\lambda =0$ generates an increase in the chiral condensate compared with
the value one would extract from the bulk of the spectrum. On the other
hand, in (\ref{cabaf}) the infinite volume limit is to be taken before the
chiral limit, and Figs.~\ref{sizfig1} and \ref{sizfig2} suggest that the
enhancement of the spectra near $\lambda =0$ weakens as the space-time
volume is increased. However, the limited statistics of the measurements
precludes a solid extrapolation of the aforementioned enhancement to
infinite volume; it is unclear whether it completely disappears as
$V\rightarrow \infty $ or whether it approaches a nontrivial limit.

Divergences in the quenched chiral condensate reminiscent of the
behavior observed here have been noted in a variety of contexts, and
both of the aforementioned alternatives concerning the
$V\rightarrow \infty $ limit have been discussed. An ensemble of vortex
configurations extracted from the full $SU(2)$ lattice Yang-Mills ensemble
using an appropriate gauge fixing and projection procedure yields a
divergent chiral condensate in the chiral limit \cite{forc2}.
This phenomenon is also observed in quenched lattice calculations using
domain wall fermions\footnote{Using more conventional lattice Dirac
operators, such behavior is not detected, presumably being veiled by
lattice artefacts \cite{hatep}.} \cite{chen}. The main effect in this case
is due to exact Dirac zero modes resulting from a nontrivial
topological charge of the gauge configurations; this effect vanishes for
$V\rightarrow \infty $ as has been argued at the end of section \ref{qsec}.
On the other hand, this effect may also hide more subtle divergences
which persist in the infinite volume limit. Such more persistent
divergences have been argued to appear both in quenched chiral
perturbation theory \cite{chipd} and in instanton models \cite{shatep}.
In the latter models, the strength of the divergence strongly decreases
as one leaves the dilute instanton gas limit and allows the associated
quark zero modes to interact strongly. Note that, in contrast to
the instanton picture, the present vortex picture does not possess an
obvious dilute limit with weakly interacting quark quasi-zero modes. In
view of this, the observation of a persistent divergence in the Dirac
spectrum as $V\rightarrow \infty $ in the instanton model does not
straightforwardly imply the same qualitative effect in the vortex picture.
More detailed information on this question may result from a
consideration of the space-time structure of low-lying quark modes
in the vortex model; such an investigation lies beyond the scope
of the present work. Note also that, in unquenched calculations,
the spectral density near zero eigenvalue will be dynamically suppressed
due to the weighting with the determinant of the Dirac operator.
Nevertheless, the aforementioned instanton model calculations still
yield divergent behavior at $\lambda =0$, which however disappears as
$m\rightarrow 0$ such as to yield a finite chiral condensate in the
chiral limit.

If the Dirac spectrum retains divergent behavior at $\lambda =0$ even
in the infinite volume limit, then the issue of the phenomenological
relevance of this behavior becomes a subtle quantitative question. 
In view of (\ref{cabaf}), for sufficiently large quark masses the
extra contribution to the chiral condensate resulting from the
divergence at $\lambda =0$ becomes negligible and the condensate
is dominated by the bulk spectral density. The question then remains
whether realistic quark masses are large or small
in this sense. While this caveat must be kept in mind, further below,
numerical values for the chiral condensate will be quoted which
correspond to the $\lambda =0$ extrapolation of the bulk of the
spectral density, disregarding the enhancement of the spectrum very
near $\lambda =0$.

Turning now to the properties of the bulk of the spectrum in more detail,
Fig.~\ref{sizfig1} displays results for the ``random $\Sigma $''
construction of the gauge field with either minimal or random monopoles,
respectively. In these cases, the bulk value evidently has already
converged well to the large volume limit at a spatial extension of the
universe corresponding to 4 lattice spacings, i.e. $4a=1.56\, $fm.
In Fig.~\ref{sizfig2}, exhibiting results for the ``smooth $\Sigma $''
construction with minimal monopoles, the convergence is not quite as fast;
there is still a discrepancy of the order of $10\, $\% between the universe
with spatial extension $1.56\, $fm and the universe with spatial extension
$1.95\, $fm, the results in the latter case being indistinguishable from
those on larger lattices\footnote{Note that the approach to the infinite
volume limit is not monotonous. This is presumably connected to the
following subtlety: In the case of a lattice with an odd number of lattice
sites in any one of the space-time dimensions, the space of quark
wave functions used here contains a mode which is not annihilated
by the derivative operator, but whose image after acting with the
derivative operator has no more overlap with any quark wave function.
Consequently, the free Dirac matrix then contains corresponding spurious zero
eigenvalues; as a result, the free Dirac equation is only solved correctly
if one uses an even number of lattice sites in all directions. In the
presence of an interaction with the gauge field, the aforementioned
problem of vanishing overlap largely disappears, as is evidenced by
the similarity of the spectra obtained for the cases of either all even
or some odd numbers of lattice sites in the four space-time directions;
however, the non-monotonous approach to the infinite volume limit is
presumably a vestige of the subtle difference between these two
cases.}. The finite size effects were also considered in
a number of other cases: Random $\Sigma $ with random monopoles on
$n^3 \times 1$ lattices\footnote{The reader is reminded that, in the
special case of a space-time of extension $a$ in the time direction, a
subdivision of the lattice in that direction is introduced when defining
the quark basis, in order to accomodate both of the lowest Matsubara modes,
cf.~section~\ref{qsec}.}, random $\Sigma $ with minimal monopoles on
$n^4 $ lattices and on $n^3 \times 1$ lattices, and smooth $\Sigma $ with
minimal monopoles on $n^4 $ lattices. All these cases displayed a convergence
to the large volume limit comparable to or better than the ones exhibited
here explicitly. Also in the smooth $\Sigma $ case with random monopoles,
the bulk of the spectrum on a $3^4 $ lattice turns out to be 
indistinguishable from the one on a $4^4 $ lattice.

\begin{figure}[h]
\centerline{
\epsfysize=9cm
\epsffile{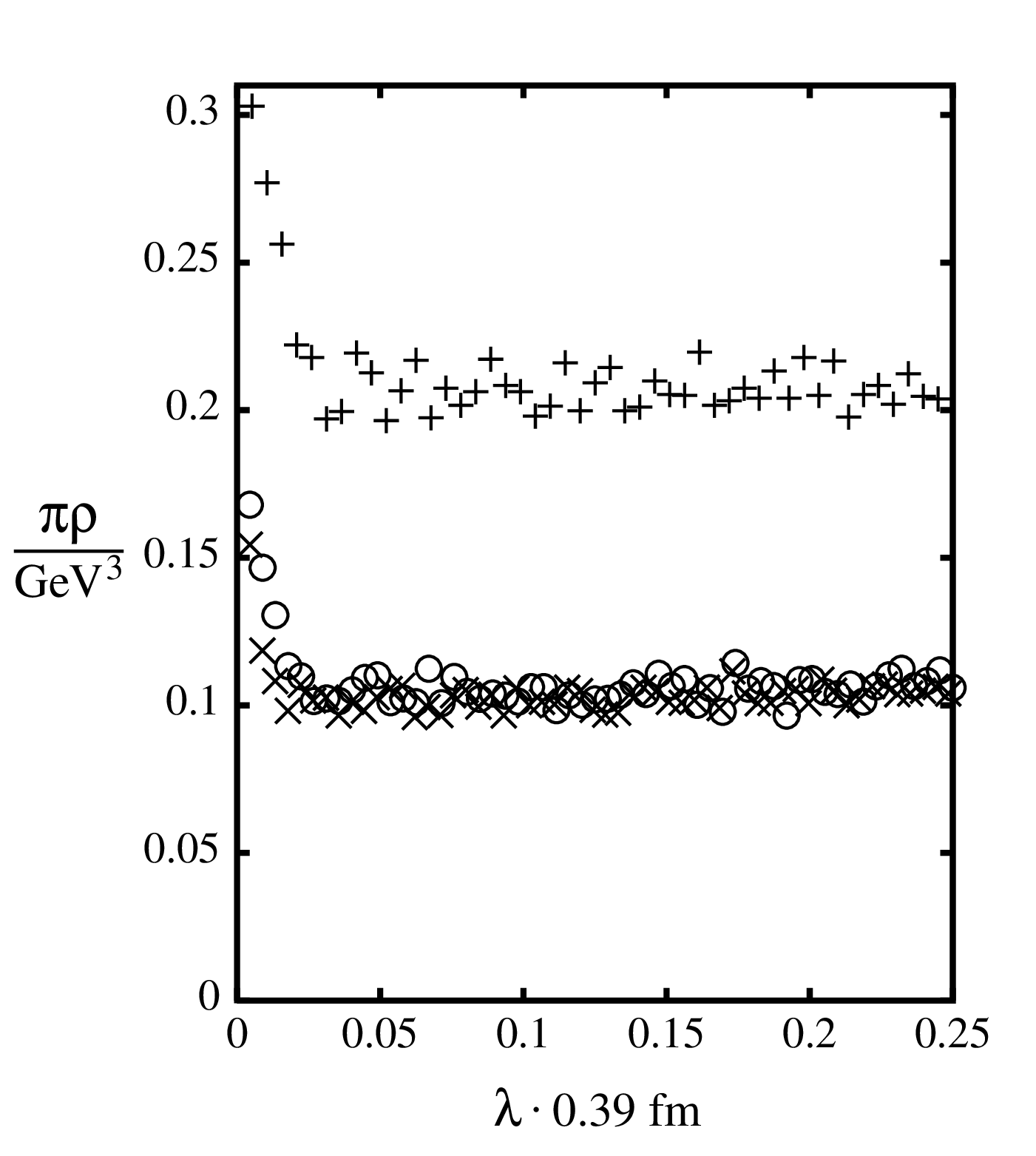}
\hspace{-0.3cm}
\epsfysize=9cm
\epsffile{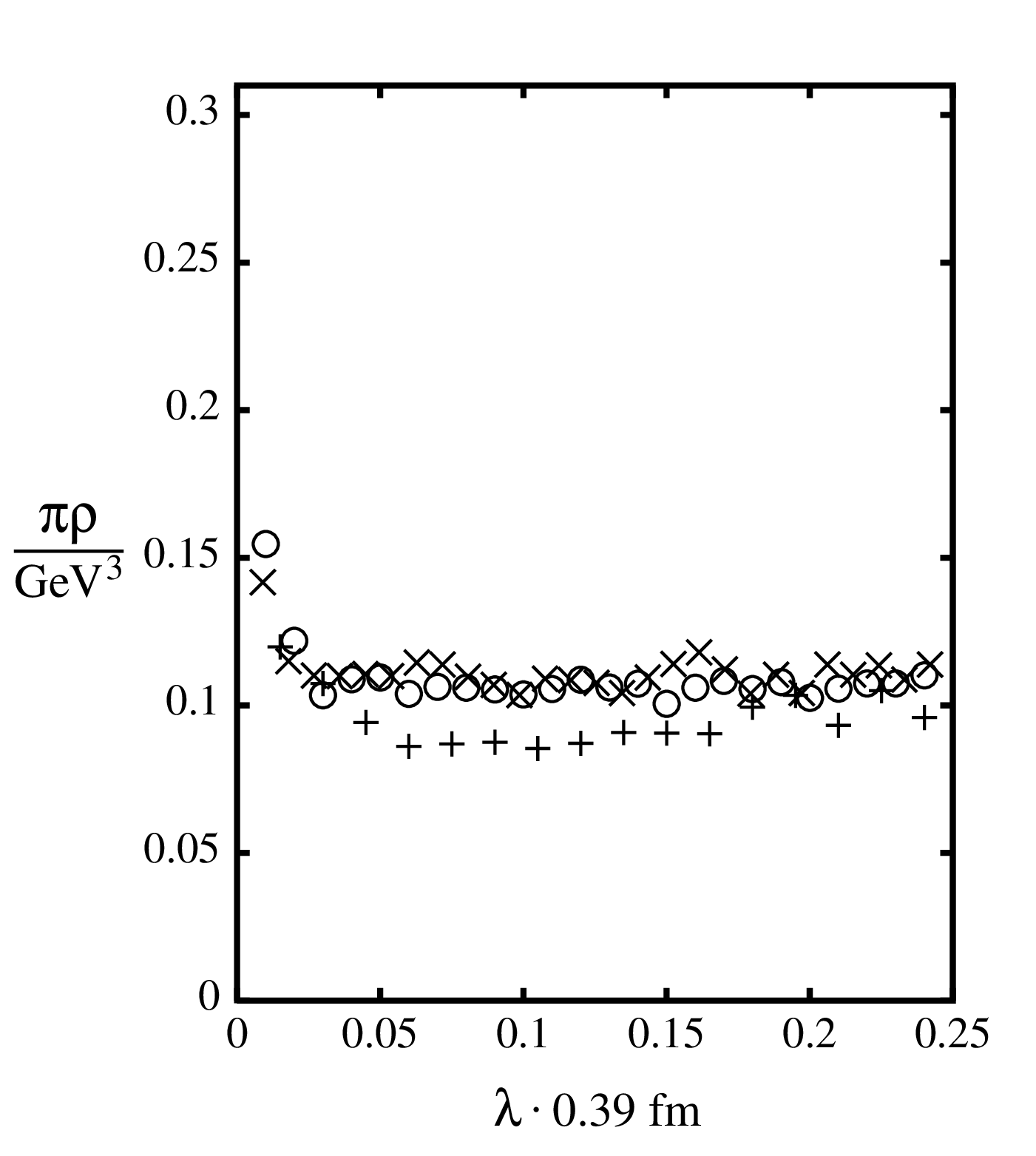}
}
\caption{Dependence of Dirac spectral density $\rho $ on temperature $T$,
at a spatial extension of the lattice of $4a=1.56\, $fm.
Left: Gauge field constructed using random monopoles and
random $\Sigma $; data symbolized by $\times $ corresponds to $T=0$,
circles correspond to $T=0.83\, T_c $, and crosses to $T=1.67\, T_c $.
Right: Gauge field constructed using minimal monopoles and random
$\Sigma $; data symbolized by $\times $ corresponds to $T=0$,
circles correspond to $T=0.83\, T_c $, and crosses to $T=1.67\, T_c $.}
\label{tfig1}
\end{figure}

\begin{figure}[h]
\centerline{
\epsfysize=9cm
\epsffile{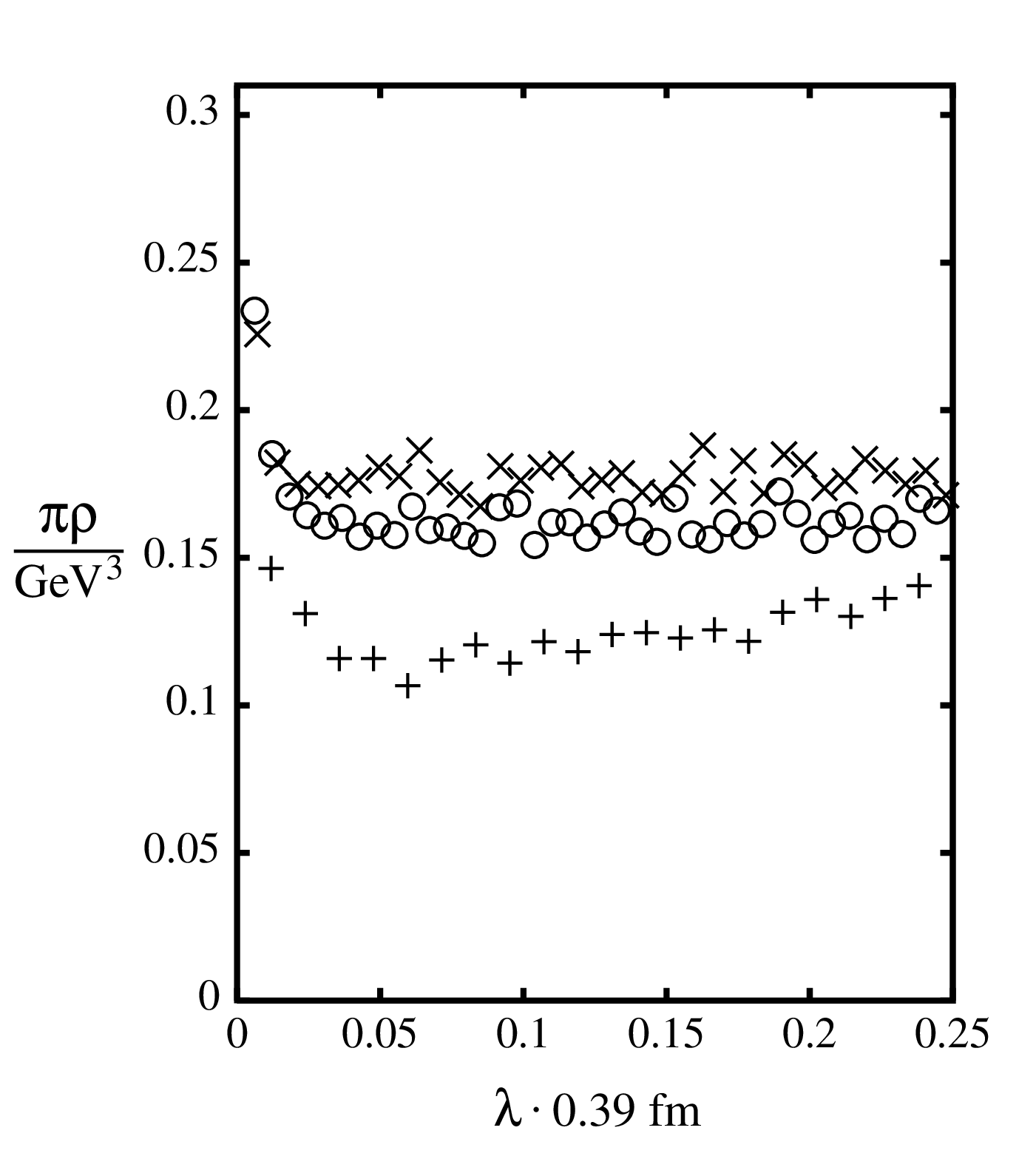}
\hspace{-0.3cm}
\epsfysize=9cm
\epsffile{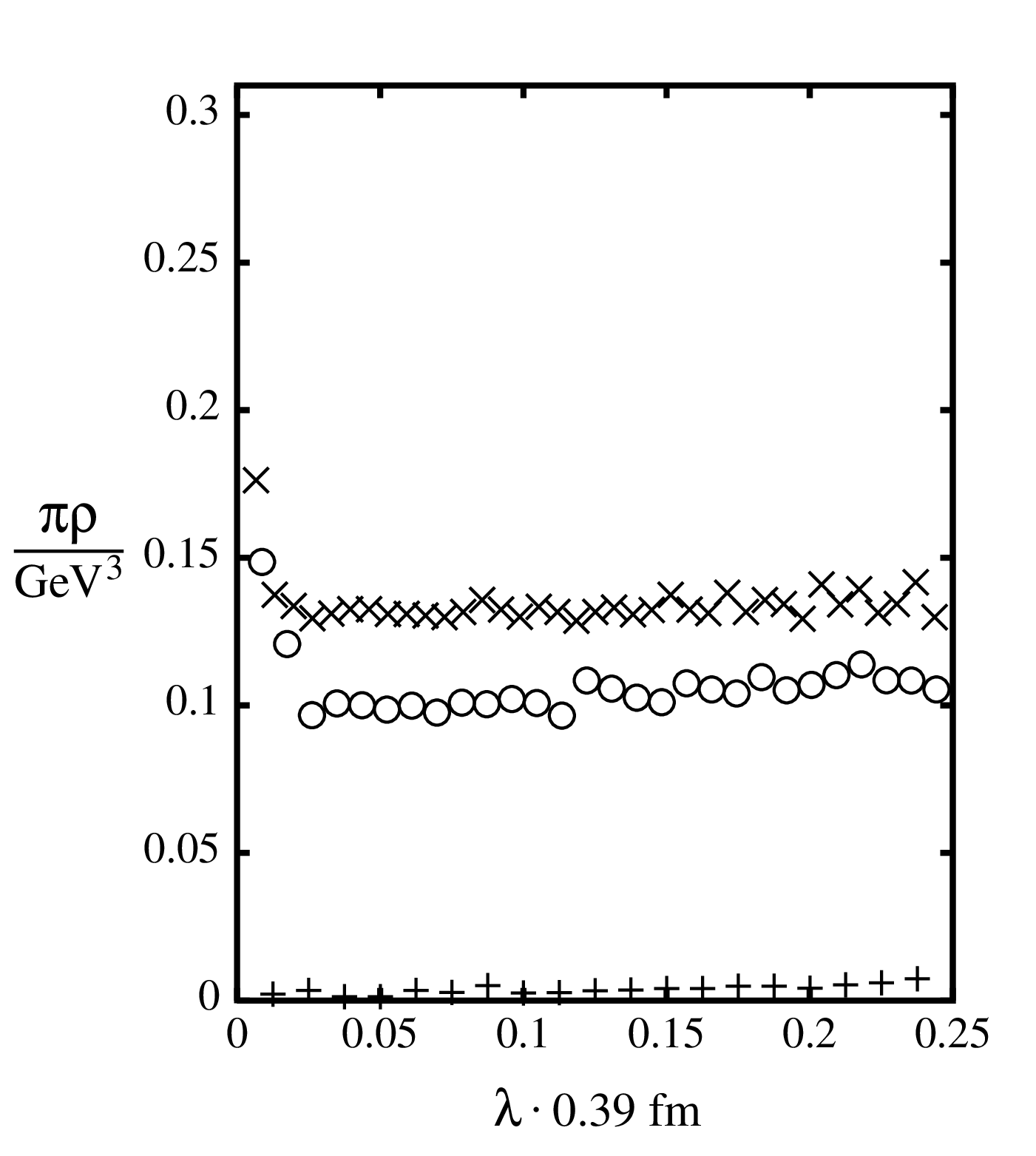}
}
\caption{Dependence of Dirac spectral density $\rho $ on temperature $T$,
at a spatial extension of the lattice of $4a=1.56\, $fm.
Left: Gauge field constructed using random monopoles and
smooth $\Sigma $; data symbolized by $\times $ corresponds to $T=0$,
circles correspond to $T=0.83\, T_c $, and crosses to $T=1.67\, T_c $.
Right: Gauge field constructed using minimal monopoles and smooth
$\Sigma $; data symbolized by $\times $ corresponds to $T=0$,
circles correspond to $T=0.83\, T_c $, and crosses to $T=1.67\, T_c $.}
\label{tfig2}
\end{figure}

\begin{figure}[h]
\centerline{
\epsfysize=9cm
\epsffile{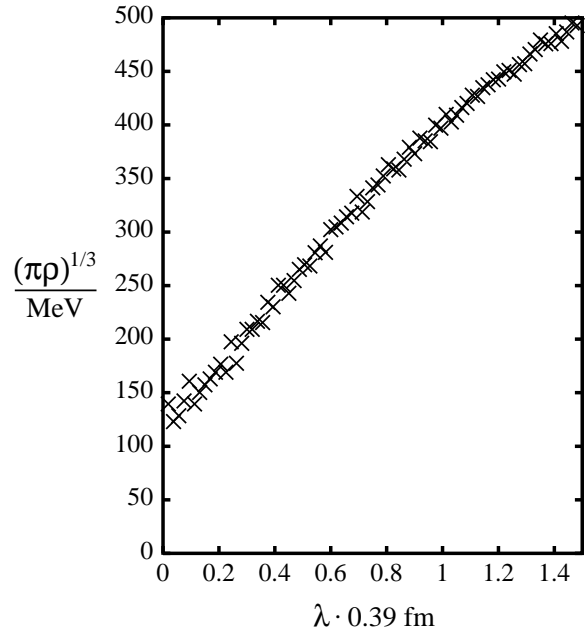}
}
\caption{Detailed plot of the (third root of the) Dirac spectral density
$\rho $ obtained at a temperature of $T=1.67\, T_c $ for a spatial
extension of the lattice of $4a=1.56\, $fm using the minimal monopole,
smooth $\Sigma $ construction of the gauge field. The chiral condensate does
not vanish in the deconfined phase, but is sharply reduced.}
\label{tfig3}
\end{figure}

In the following, Dirac spectra will be shown for different choices of the
gauge field construction and different temperatures at a spatial extension
of the lattice of $4a=1.56\, $fm. While one should be aware that this
extension in some cases still allows for finite size effects of the order
of $10\, $\%, it does not suffer from the aforementioned subtlety
(cf.~footnote) arising for lattices with odd numbers of lattice sites in
at least one of the space-time directions, and it also allows for superior
statistics compared with extensions corresponding to the next even multiple
of the lattice spacing, $6a=2.34\, $fm. The results are depicted in
Figs.~\ref{tfig1}-\ref{tfig3}, and also summarized in Table~\ref{table1}.

\begin{table}[h]

\[ 
\begin{array}{|c||c|c|c|c|}
\hline
& \multicolumn{4}{c|}{-\langle \bar{q} q\rangle / \mbox{GeV}^{3} } \\
\cline{2-5}
T/T_c & \mbox{minim. monop.} &
\mbox{minim. monop.} &
\mbox{random monop.} &
\mbox{random monop.} \\
& \mbox{smooth } \Sigma &
\mbox{random } \Sigma &
\mbox{smooth } \Sigma &
\mbox{random } \Sigma \\
\hline\hline
0 & 0.129 & 0.108 & 0.175 & 0.100 \\ \hline
0.83 & 0.094 & 0.100 & 0.157 & 0.097 \\ \hline
1.1 & 0.059 & \mbox{---} & \mbox{---} & \mbox{---} \\ \hline
1.4 & 0.0033 & \mbox{---} & \mbox{---} & \mbox{---} \\ \hline
1.67 & 0.0015 & 0.078 & 0.103 & 0.203 \\ \hline
\end{array}
\]
\vspace{0.3cm}
 
\caption{Chiral condensate as a function of temperature for different
gauge field constructions on a lattice of spatial extension 1.56 fm,
obtained by extrapolating the bulk of the Dirac spectral density to
zero eigenvalue. Since the inverses of the temperatures $T=1.1\, T_c $
and $T=1.4\, T_c $ are not integer multiples of the lattice spacing $a$ in
the random surface model with curvature coefficient $c=0.24$, the
associated values of the chiral condensate were obtained by the following
interpolation procedure \cite{selprep}. The inverse of $T=1.1\, T_c $
corresponds to one lattice spacing for $c=0.14$ and to two lattice spacings
for $c=0.32$; interpolating $\langle \bar{q} q\rangle \, a^3 $ as a function
of $c$ defines the chiral condensate also at $c=0.24$ for $T=1.1\, T_c $.
The same procedure was also applied to $T=1.4\, T_c $, the inverse of which
corresponds to one lattice spacing for $c=0.2$ and to two lattice
spacings for $c=0.407$.}

\label{table1}

\end{table}

Evidently, in the confined phase, all models yield a bulk spectral
density which extrapolates to a nonzero value at $\lambda =0$,
thus signaling the spontaneous breaking of chiral symmetry.
Quantitatively\footnote{The scale in the random vortex surface model is
fixed by equating the zero-temperature string tension with
$(440\, \mbox{MeV})^{2} $, simultaneously implying a lattice spacing
of $a=0.39\, $fm in the model with curvature coefficient $c=0.24$,
cf. \cite{selprep}.}, the zero-temperature quenched chiral condensate
obtained for the different gauge field constructions varies between
$-0.1\, \mbox{GeV}^{3} $ and $-0.175\, \mbox{GeV}^{3} $. For comparison, the
value obtained in lattice Yang-Mills calculations \cite{hatep} corresponds
to $-0.09\, \mbox{GeV}^{3} $ (with the zero-temperature string tension fixed
to the same value as in the present model). In particular, both of the
random $\Sigma $ models come very close to the lattice Yang-Mills result.
However, it should be kept in mind that the actual value of
$\langle \bar{q} q\rangle $ does not have direct phenomenological
meaning, since it is not a renormalization group invariant
quantity\footnote{At this stage, the relation between the ultraviolet
cutoff scheme of the present vortex model on the one hand and, on the
other hand, standard renormalization schemes such as the ones used on
the lattice or in perturbation theory is not fixed. In principle,
matching different schemes can lead to large correction terms; this
is well known in the case of matching lattice renormalization with
standard perturbation theory schemes, where e.g.~the chiral condensate
acquires corrections of the order of $30$\% \cite{hatep}.}.
Only after multiplying with the quark mass, one obtains a physically
meaningful quantity which enters e.g.~current algebra relations. In
this sense, a measurement of $\langle \bar{q} q\rangle $ can be
viewed as representing no more than an accessory to
defining the current quark masses within the model. However, the fact that
the vortex model yields values for the chiral condensate which are not
separated from lattice values by unnaturally large factors gives rise to
the expectation that this model succeeds in capturing the relevant
physics leading to spontaneous chiral symmetry breaking in the
confined phase of the strong interaction. More stringent quantitative
tests of low-energy quark physics within the vortex model demand the
calculation of additional observables; this is deferred to future work.

Turning now to the behavior of the chiral condensate as one crosses
the finite-temperature deconfinement transition, the different models
evidently display much larger, qualitative, differences. The random
$\Sigma $, random monopole model yields a considerably enhanced
chiral condensate in the deconfined phase; the other models
show a reduction of the condensate above the critical temperature.
However, this reduction is rather slight, except in the case of the
smooth $\Sigma $, minimal monopole model. This latter case is the
only one which qualitatively agrees with lattice calculations
\cite{chen}: The quenched chiral condensate does not disappear in the
deconfined regime, but it is sharply reduced as one crosses the phase
transition, cf.~in particular Fig.~\ref{tfig3}. In view of the discussion
of section \ref{gensec}, the fact that the smoothest choice of gauge
field construction yields the most realistic physics is not unexpected.
This appears to be the model of choice. As discussed above, the fact
that it yields a value for the quenched chiral condensate in the confining
phase which is slightly enhanced compared with lattice calculations has no
direct phenomenological relevance. Alternatively, one could envisage
ultimately using a minimal monopole model which interpolates between
the smooth and random $\Sigma $ cases such as to reproduce the lattice
results even more closely.

\section{Summary}
Three fundamental nonperturbative phenomena characterize the low-energy
regime of the strong interaction: Confinement, the $U_A (1)$ flavor anomaly,
and the spontaneous breaking of chiral symmetry. While the ($SU(2)$ color)
random vortex world-surface model was shown to quantitatively describe the
first two of the aforementioned phenomena in \cite{selprep},\cite{preptop},
the present work indicates that also the third phenomenon, i.e.~the
spontaneous breaking of chiral symmetry, finds a viable description
within this model. Specifically, a gauge field construction for vortex
world-surface configurations was found which, coupled to quarks, leads to a
behavior of the quenched chiral condensate as a function of temperature in
qualitative agreement with the behavior found in lattice Yang-Mills theory.
Quantitatively, the quenched chiral condensate $\langle \bar{q} q\rangle $
differs from the lattice Yang-Mills result by a factor of the order of unity
which presumably can be absorbed into the ultraviolet regularization scheme
dependence of the chiral condensate, or, equivalently, into the definition
of the current quark mass $m$. Only the product $m\langle \bar{q} q\rangle $
is a renormalization group invariant, physical quantity which can be
tested e.g.~using current algebra techniques. Stringent quantitative
tests of low-energy quark physics via additional hadronic observables
are deferred to future work.

The main thrust of the present investigation lay in developing the technical
tools necessary for such work. In order to incorporate quark physics
into the vortex model, the Dirac operator in a vortex background had
to be constructed. This involves, as a prerequisite, casting vortex 
configurations explicitly in terms of gauge fields. Having appropriately
specified such vortex gauge fields via a Wu-Yang construction, a matrix
representation of the Dirac operator in a truncated (finite element) quark
basis was obtained which determines infrared quark propagation in the
combined vortex-quark system. The evaluation of the quenched chiral
condensate highlighted above represents a first exploratory application
of these techniques. 

With the Dirac operator at hand, one can envisage also carrying out
dynamical quark calculations within the vortex model, by reweighting
the vortex ensemble with the Dirac operator determinant. This will
substantially penalize Dirac eigenvalues of very small magnitude and
thus presumably reduce the chiral condensate to the phenomenologically
expected values around $(230\, \mbox{MeV})^3 $. Moreover, to make contact
with phenomenology, the random vortex surface model must still be
extended to $SU(3)$ color; also in the three-color case, lattice
Yang-Mills measurements \cite{fabdo3} indicate that vortex degrees
of freedom dominate the infrared, confining sector of the strong
interaction.

\section*{Acknowledgments}
The author acknowledges fruitful and informative discussions with
C.~Alexandrou, R.~Alkofer, P.~van~Baal, R.~Bertle, J.~Bloch, M.~Faber,
P.~de~Forcrand, L.~Gamberg, T.~Kov\'acs, H.~Reinhardt, P.~Watson and
H.~Weigel.

\end{document}